\documentclass[11pt,reqno]{amsart}

\usepackage[utf8]{inputenc}

\usepackage{geometry}
\usepackage{amsmath,amsthm,amssymb}
\usepackage{graphicx}
\usepackage{mathabx}
\usepackage{color}
\usepackage{cite}
 
\newcommand{\bPsi}{\bf\Psi} 
\newcommand{\modei}{{a_+}}
\newcommand{\modeii}{{a_-}}
\newcommand{\modem}{{a_m}}
\newcommand{\moden}{{a_n}}
\newcommand{\coup}{{g}} 
\newcommand{\br}{{\bf r}}

\newcommand{\bQ}{{\bf Q}}
\newcommand{\bq}{{\bf q}}

\newcommand{\bp}{{\bf p}}

\newcommand{\bu}{{\bf u}}
\newcommand{\bb}{{\bf b}}

\newcommand{\bv}{{\bf v}}

\newcommand{\bk}{{\bf k}}

\newcommand{\bX}{{\bf X}}

\newcommand{\bx}{{\bf x}}
\newcommand{\by}{{\bf y}}

\newcommand{\hbk}{{\hat{\bk}}}

\newcommand{\dd}{{{d}}}

\newcommand{\im}{\mathrm{Im} \, }
\newcommand{\re}{\mathrm{Re} \, }

\title{Collective Spontaneous emission in random media}

\author{Joseph Kraisler}
\address{Department of Applied Physics and Applied Mathematics, Columbia University, New York, NY 10027}
\email{jek2199@columbia.edu}

\author{John C. Schotland}
\address{Department of Mathematics, Yale University, New Haven, CT  06511}
\email{john.schotland@yale.edu}

\date{\today}

\begin{document}

\begin{abstract}
We consider the theory of spontaneous emission for a random medium of stationary two-level atoms. We investigate the dynamics of the field and atomic probability amplitudes for a one-photon state of the system. At long times and large distances, we show that the corresponding average probability densities can be determined from the solutions to a pair of kinetic equations.  
\end{abstract}

\maketitle

\section{Introduction}

The quantum theory of light-matter interactions has historically been concerned with systems consisting of a small number of atoms~\cite{Mandel_1995}. To some extent, this situation is due to the early emphasis on such systems in atomic physics. However, the recent focus on cold atom systems~\cite{Haroche_2006,Gardiner_2015}, waveguide quantum electrodynamics~\cite{Liao_2016,Roy_2017}, and semiconductor quantum optics~\cite{Kira_2011}, has served to stimulate research on quantum many-body problems. Progress in this direction can be expected to lead to significant advances in controlling quantum systems, with applications to quantum simulations, quantum information processing, and precision measurements~\cite{Kimble_2008,Riedmatten_2008,Bloch_2012}.


Perhaps the simplest many-body problem in quantum optics arises in a system of two-level atoms interacting with a single photon. Suppose that one of the atoms is initially in its excited state and there are no photons present in the field. The atom can then decay by spontaneous emission, thereby transferring its excitation to the field. The resulting photon can then excite the remaining atoms, which likewise decay. This process, which is referred to as collective or cooperative emission,  results in the transmission of light through the system. Two regimes are usually distinguished, depending on the wavelength and the size of the system: superradiance and radiation trapping. In single-photon superradiance, certain states decay much faster than the single-atom decay rate. Alternatively, there is very slow decay, and the states are said to be trapped. Moreover, in contrast to single-atom spontaneous emission, where the Lamb shift is divergent, the Lamb shift can be finite in single-photon superradiance. 

The theory of collective emission has been considered from several points of view. One approach is based on a Hamiltonian describing the atoms, the optical field and their interaction. Eliminating the optical field yields an effective Hamiltonian for the atomic degrees of freedom~\cite{Lehmberg_1970,James_1993,Bienaime_2014,Zhu_2016,Mirza_2016}. A master equation can then be derived, and has been shown to describe quantum effects in light scattering. However, the computational cost of this procedure, which scales exponentially with the number of atoms, limits its utility to systems consisting of a small number of atoms.
An alternative approach, which makes use of the eigenstates and corresponding eigenvalues of the effective Hamiltonian, can be employed to describe the dynamics of the system~\cite{Friedberg_1972,Friedberg_1973,Friedberg_2008,Scully_2006,Svidzinsky_2008,Svidzinsky_2010,Bienaime_2012}. This method is especially fruitful in the setting of single-photon superradiance, where analytical expressions for the collective decay rate have been obtained for dense atomic gases.

In this paper, we consider the problem of collective emission for a random medium of two-level atoms. 
In this setting, we investigate the dynamics of the field and atomic probability amplitudes for a one-photon state of the system. At long times and large distances, we find that the corresponding average probability densities can be determined from the solutions to a pair of kinetic equations. There are several novel mathematical aspects of our work. We employ a real-space quantization procedure for the optical field. In contrast, quantization of the field is normally carried out in terms of Fourier modes. The advantage of the real-space approach is that it allows the field and atomic degrees of freedom to be treated on an equal footing.
Moreover, the field and atomic probability amplitudes obey a system of \emph{nonlocal} partial differential equations with random coefficients. Using this result, we show that the average Wigner transform of the amplitudes obeys a kinetic equation, whose diffusion limit is extracted. Here the average over the random medium is carried out by means of a multiscale asymptotic expansion in a suitable high-frequency limit~\cite{Ryzhik_1996,Bal_2005,Caze_2015,Carminati_2020}. 

This paper is organized as follows. In section~\ref{model} we introduce the model we study, carry out the real-space quantization of the optical and atomic fields, and derive the equations obeyed by the atomic and one-photon amplitudes. These equations are studied in section~\ref{single-atom} for the case of a single atom, where we recover the Wigner-Weisskopf theory of spontaneous emission, and in section~\ref{constant-density} for the case of a medium of constant density. Random media are introduced in sections~\ref{eigenstates}, where the average behavior of energy eigenstates is established. A related approach leads to the derivation of kinetic equations. The paper concludes with a discussion of our results in section~\ref{discussion}. The technical details of certain calculations are presented in the appendices.

\section{Model}
\label{model}
We consider the following model for the interaction between a quantized field and a system of  two-level atoms~\cite{Dicke_1954,Gross_1982}. The atoms are taken to be stationary and sufficiently well separated that interatomic interactions can be neglected. 
The overall system is described by the Hamiltonian $H=H_F + H_A + H_I$. 
The Hamiltonian of the field is of the form
\begin{align}
H_F = \int\frac{d^3 k}{(2\pi)^3} \hbar\omega_\bk\, a^{\dagger}_{\bk}a_{\bk} \  ,
\label{HF}
\end{align}
where we have neglected the zero-point energy and for simplicity have adopted a scalar theory of the electromagnetic field. Here $\omega_\bk=c\vert\bk\vert$ is the frequency of the field mode with wave vector $\bk$ and $a^{\dagger}_{\bk}$ ($a_{\bk}$) is the corresponding creation (annihilation) operator. The operators $a_{\bk}$ and $a^{\dagger}_{\bk}$ obey the commutation relations
\begin{align}
\label{commutator}
[a_{\bk},a^{\dagger}_{\bk'}]&=\delta(\bk-\bk') \ , \\
[a_{\bk},a_{\bk'}]&=0 \ . 
\end{align}
The Hamiltonian of the atoms is given by
\begin{align}
H_A = \sum_{j}\hbar\Omega\sigma_j^{\dagger}\sigma_j \  ,
 \label{HA}
\end{align}
where $\Omega$ is the resonance frequency of each atom and $\sigma_j^\dag$ ($\sigma_j$) is the raising (lowering) operator of the $j$th atom. The operators $\sigma_j$ and  $\sigma^{\dagger}_j$ obey the anticommutation relations
\begin{align}
\{\sigma_j,\sigma^{\dagger}_{j'}\}&=\delta_{jj'} \ , \\
\{\sigma_j,\sigma_{j'}\}&=0 \ .
\end{align}
The interaction between the field and the atoms is governed by the Hamiltonian
 \begin{align}
H_I = \sum_{j}\int\frac{d^3 k}{(2\pi)^3} \hbar g_{\bk}\left(a_{\bk}+a^{\dagger}_{\bk}\right)\left(e^{i\bk\cdot\bx_j}\sigma_j+e^{-i\bk\cdot\bx_j}\sigma^{\dagger}_j\right) \  ,
\label{HI}
\end{align}
where $g_{\bk}$ is the field-atom coupling and $\bx_j$ is the position of the $j$th atom.

In order to treat the atoms and the field on the same footing, it is useful to introduce a real-space representation of the Hamiltonian (\ref{HF}). To this end, we define the operator $\phi(\bx)$ as the Fourier transform of $a_\bk$:
\begin{align}
\phi(\bx) &=\int\frac{d^3 k}{(2\pi)^{3/2}} e^{i\bk\cdot\bx}a_{\bk} \ . 
\label{eq:b3}
\end{align}
Making use of (\ref{commutator}) we find that $\phi$ is a Bose field which obeys the commutation relations
\begin{align}
\label{commutation}
[\phi(\bx),\phi^{\dagger}(\bx')]&=\delta(\bx-\bx') \ , \\
[\phi(\bx),\phi(\bx')]&=0 \ .
\end{align}
It follows immediately that $H_F$ becomes
\begin{align}
H_F =   \hbar c \int d^3 x (-\Delta)^{1/2}\phi^{\dagger}(\bx)\phi(\bx) \ .
\end{align}
Here the operator $(-\Delta)^{1/2}$ is defined by the Fourier integral
\begin{align}
(-\Delta)^{1/2}f(\bx) &= \int\frac{d^3 k}{(2\pi)^3} e^{i\bk\cdot\bx}\vert\bk\vert\tilde{f}(\bk) \ , \\
\tilde{f}(\bk) &= \int d^3 x  e^{-i\bk\cdot\bx}f(\bx) \ .
\end{align}
We note that $(-\Delta)^{1/2}$ has the non-local spatial representation
\begin{align}
(-\Delta)^{1/2}f(\bx) = \frac{1}{\pi^2}\int d^3 y\, \frac{f(\bx)-f(\by)}{\vert\bx-\by\vert^4} \ .
\end{align}
We note that real-space quantization has proven to be a powerful tool for one-dimensional systems in the setting of waveguide quantum electrodynamics~\cite{Shen_2005}.

To facilitate the treatment of random media, it will prove convenient to introduce a continuum model of the atomic degrees of freedom. The atomic Hamiltonian then becomes
\begin{align}
H_A = \hbar\Omega\int d^3x \rho(\bx)\sigma^{\dagger}(\bx)\sigma(\bx) \ ,
\end{align}
where $\rho$ is the number density of atoms. In addition, the operators $\sigma_j$ are replaced by a Fermi field $\sigma$ which obeys the anticommutation relations
\begin{align}
\label{anticommutation}
\{\sigma(\bx),\sigma^{\dagger}(\bx)\}&=\frac{1}{\rho(\bx)}\delta(\bx-\bx') \ , \\
\{\sigma(\bx),\sigma(\bx)\}&=0 \ .
\end{align}
We find that the interaction Hamiltonian is given by
\begin{align}
H_I = \hbar \coup\int d^3x \rho(\bx)\left(\phi(\bx) + \phi^\dagger(\bx)\right) \left(\sigma(\bx)+\sigma^{\dagger}(\bx) \right) \ ,
\end{align}
where we have made the Markovian approximation $g_\bk = g$ for all $\bk$, so that the atom-field coupling is frequency independent. We also impose the rotating wave approximation (RWA), in which we neglect the rapidly oscillating terms $\phi^{\dagger}\sigma^{\dagger}$ and $\phi\sigma$. The total Hamiltonian thus becomes
\begin{align}
\label{Hamiltonian}
H = \hbar\int d^3x \left[c(-\Delta)^{1/2}\phi^{\dagger}(\bx)\phi(\bx) +\Omega\rho(\bx)\sigma^{\dagger}(\bx)\sigma(\bx) + \coup\rho(\bx)\left(\phi^{\dagger}(\bx)\sigma(\bx)+\phi(\bx)\sigma^{\dagger}(\bx) \right)\right] \ ,
\end{align}
which is the model we will investigate for the remainder of this paper.

We suppose that the system is in a one-photon state of the form
\begin{align}
\label{one_photon_state}
\vert\Psi\rangle = \int d^3 x \left[\psi(\bx,t)\phi^{\dagger}(\bx) + \rho(\bx)a(\bx,t)\sigma^{\dagger}(\bx)\right]\vert 0\rangle \ ,
\end{align}
where $\vert 0\rangle$ is the combined vacuum state of the field and the ground state of the atoms. Here $a(\bx,t)$ denotes the probability amplitude for exciting an atom at the point $\bx$ at time $t$ and $\psi(\bx,t)$ is the  amplitude for creating a photon. The state
$\vert\Psi\rangle$ is the most general one-photon state that is consistent with the RWA. In addition, $\vert\Psi\rangle$ is normalized so that $\langle \Psi\vert\Psi\rangle = 1$. It follows from (\ref{commutation}) and (\ref{anticommutation}) that the amplitudes obey the normalization condition
\begin{align}
\int d^3x \left(\vert\psi(\bx,t)\vert^2 + \rho(\bx)\vert a(\bx,t)\vert^2\right) = 1
\ .
\end{align}

The dynamics of $\vert\Psi\rangle$ is governed by the Schrodinger equation
\begin{align} 
\label{eq:schr}
i\hbar\partial_t\vert\Psi\rangle = H\vert\Psi\rangle \ .
\end{align}
Projecting onto the states $\phi^{\dagger}(\bx)\vert 0\rangle$ and $\sigma^{\dagger}(\bx)\vert 0\rangle$ and making use of (\ref{commutation}) and (\ref{anticommutation}), we arrive at the following system of equations obeyed by $a$ and $\psi$:
\begin{align}
\label{eq:b4}
i\partial_t\psi & = c(-\Delta)^{1/2}\psi + \coup\rho(\bx) a \ , \\
i\rho(\bx)\partial_t a & = \coup\rho(\bx)\psi + \Omega\rho(\bx) a \ .
\label{eq:b5}
\end{align}
The details of the derivation are given in Appendix A. The overall factors of $\rho(\bx)$ in (\ref{eq:b5}) will be cancelled as necessary.


\section{Single Atom}
\label{single-atom}

In this section we consider the problem of spontaneous emission by a single atom. We assume that the atom is located at the origin and put $\rho(\bx)=\delta(\bx)$. We also assume that the atom is initially in its excited state and that there are no photons present in the field. We thus impose the initial conditions $a(\bx,0)=1$ and $\psi(\bx,0)=0$. Taking the Laplace transform in $t$ and the Fourier transform in $\bx$ of (\ref{eq:b4}) and (\ref{eq:b5}), and applying the initial conditions gives
\begin{align}
\label{eq:c1}
iz\tilde{\psi}(\bk,z) & = c\vert\bk\vert\tilde{\psi}(\bk,z) + \coup a(0,z) \ , \\
i(z a(0,z)-1) & = \coup\psi(0,z) + \Omega a(0,z) \ .
\label{eq:c2} 
\end{align}
Here we have defined the Laplace transform by
\begin{align}
f(z)=\int_0^{\infty}dt e^{-zt}f(t) \ ,
\end{align}
and we denote a function and its Laplace transform by the same symbol.
Solving the above equations by making use of the relation
\begin{equation}
\psi(0,z) = \int \frac{d^3k}{(2\pi)^3}\tilde\psi(\bk,z) \ ,
\end{equation}
leads to an expression for $a(0,z)$ of the form
\begin{align}
\label{eq:a(0,z)}
a(0,z)=\frac{1}{z+i\Omega-i\Sigma(z)} \ ,
\end{align}
where $\Sigma$ is defined by 
\begin{align}
\Sigma(z)= g^2\displaystyle\int\frac{d^3k}{(2\pi)^3}\frac{1}{c\vert\bk\vert-iz-i\epsilon} \ ,
\end{align}
where $\epsilon \to 0^+$. Inverting the the Laplace transform in (\ref{eq:a(0,z)}), we obtain
\begin{equation}
\label{laplace_inv}
a(0,t) = \int \frac{dz}{2\pi i} \frac{e^{zt}}{z+i\Omega-i\Sigma(z)} \ .
\end{equation}
In order to carry out the integral (\ref{laplace_inv}), we make the pole approximation 
in which we evaluate $\Sigma$ near resonance. That is, we replace $\Sigma(z)$ with $\Sigma(-i\Omega)$. In addition, we split $\Sigma(-i\Omega)$ into its real and imaginary parts:
\begin{align} 
\label{re}
\re\Sigma(-i\Omega) &=\delta\omega \ ,\\
\label{im}
\im\Sigma(-i\Omega) &=\Gamma/2 \ ,
\end{align}
which defines $\delta\omega$ and $\Gamma$.
By making use of the identity
\begin{align}
\frac{1}{c\vert\bk\vert-\Omega-i\epsilon} = P \frac{1}{c\vert\bk\vert-\Omega}+i\pi\delta(c\vert\bk\vert-\Omega) \ , 
\end{align}
where $P$ denotes the principal value, we find that $\Gamma$ is given by
\begin{align}
\label{def_Gamma}
\Gamma &= 2g^2\pi\displaystyle\int\frac{d^3 k}{(2\pi)^3}\,\delta(c\vert\bk\vert-\Omega)\\
&=\frac{g^2\Omega^2}{\pi c^3} \ .
\end{align}
We also obtain
\begin{equation}
\label{def_lamb}
\delta\omega = \frac{g^2}{2\pi^2} \int_0^{2\pi/\Lambda} \frac{k^2 dk}{ck-\Omega} \ ,
\end{equation}
where we have introduced a high-frequency cutoff to regularize the divergent integral. Finally, making use of (\ref{laplace_inv}), (\ref{re}) and (\ref{im}), we find that $a$ is given by
\begin{align}
a(0,t) &=e^{-i(\Omega-\delta\omega)t}e^{-\Gamma t/2} \ . 
\end{align}
We immediately see that the probability the atom decays is exponentially decreasing:
\begin{align}
\vert a(0,t)\vert^2 &= e^{-\Gamma t} \ .
\end{align}
We note that the decay rate $\Gamma$ agrees with Wigner-Weisskopf theory formulated within scalar electrodynamics and that $\delta\omega$ is the corresponding Lamb shift.

Next we determine the behavior of the amplitude $\psi$. Making use of  (\ref{eq:c1}), (\ref{eq:a(0,z)}) and inverting the Laplace transform, we find that
\begin{align}
\tilde\psi(\bk,t) = \int \frac{dz}{2\pi i} \frac{g e^{zt}}{\left(iz-c|\bk|\right)\left(z+i\Omega -i\Sigma(z)\right)} \ .
\end{align}
Carrying out the above integral in the pole approximation, we obtain
\begin{align}
\tilde\psi(\bk,t) = \frac{g}{c|\bk| - (\Omega-\delta\omega)+i\Gamma/2}\left(e^{-ict|\bk|} - e^{-\Gamma t/2} e^{-i(\Omega-\delta\omega)t}\right) \ .
\end{align}
At long times ($\Gamma t \gg 1$), we see that the one-photon probability density is given by
\begin{align}
|\tilde\psi(\bk,t)|^2 = \frac{|g|^2}{\left[c|\bk| - (\Omega-\delta\omega)\right]^2+\Gamma^2/4} \ ,
\end{align}
which has the form of a Lorentzian spectral line.

\section{Constant Density Problem}
\label{constant-density}
In this section we consider the problem of emission and absorption of one photon interacting with a collection of atoms with constant number density $\rho_0$. We will start with (\ref{eq:b4}) and (\ref{eq:b5}) and $\rho(\bx)$ equal to $\rho_0$. That is
\begin{align}
\label{eq_psi}
i\partial_t\psi & = c(-\Delta)^{1/2}\psi + \coup\rho_0 a \ , \\
i\partial_t a & = \coup\psi + \Omega a \, ,
\label{eq_a}
\end{align}
where we have cancelled the density from (\ref{eq:b4}). Defining the vector quantity $\bPsi(\bx,t)$ as
\begin{align}
    \bPsi(\bx,t)=\begin{bmatrix} \psi(\bx,t) \\ \sqrt{\rho_0}a(\bx,t)\end{bmatrix}\, ,
\end{align}
then the previous system becomes
\begin{align}\label{eq:constdensity}
    i\partial_t \bPsi = A\bPsi\, ,
\end{align}
where
\begin{align} \label{eq:matrix}
    A(\bx) &= \begin{bmatrix}c(-\Delta)^{1/2} & \coup\sqrt{\rho_0}\\ \coup\sqrt{\rho_0} & \Omega \end{bmatrix}.
\end{align}
Taking the Fourier transform of (\ref{eq:constdensity}), we arrive at the system of ordinary differential equations
\begin{align}
    i\partial_t \hat{\bPsi} = \hat{A}\hat{\bPsi}\, ,
\end{align}
where
\begin{align}\label{eq:Fourierconstdensity}
    \hat{\bPsi}(\bk,t) & =\int d^3 x e^{-i\bk\cdot\bx}\bPsi(\bx,t)\, ,
\end{align}
and
\begin{align} \label{eq:fouriermatrix}
    \hat{A}(\bk) &= \begin{bmatrix}c\vert\bk\vert & \coup\sqrt{\rho_0}\\ \coup\sqrt{\rho_0} & \Omega \end{bmatrix}.
\end{align}
The eigenvalues and eigenvectors of $\hat{A}$ are given by
\begin{align}
    \lambda_{\pm}(\bk) &=\frac{(c\vert\bk\vert+\Omega)\pm\sqrt{(c\vert\bk\vert-\Omega)^2+4\coup^2\rho_0}}{2} \, ,\\
    \bv_{\pm}(\bk) &= \frac{1}{\sqrt{(\lambda_{\pm}-\Omega)^2+\coup^2\rho_0}}\begin{bmatrix}
                \lambda_{\pm}-\Omega\\
                \coup\sqrt{\rho_0}
                \end{bmatrix}.
\end{align}
The solution to ~(\ref{eq:Fourierconstdensity}) is given by
\begin{align} \label{eq:constsolution}
    \hat{\bPsi}(\bk,t)= C_+(\bk)e^{-i\lambda_{+}t}\bv_{+}(\bk)+C_-(\bk)e^{-i\lambda_{-}t}\bv_{-}(\bk).
\end{align}
Solving for the coefficients $C_{\pm}(\bk)$ we find 
\begin{align} 
    C_{+}(\bk)&=\frac{(\hat{\psi_0}g-\hat{a_0}(\lambda_{-}-\Omega))\sqrt{(\lambda_{+}-\Omega)^2+g^2\rho_0}}{g(\lambda_+-\lambda-)}\label{eq:coefficient1} \ , \\
    C_{-}(\bk)&=\frac{(\hat{a_0}(\lambda_{+}-\Omega)-g\hat{\psi_0})\sqrt{(\lambda_{-}-\Omega)^2+g^2\rho_0}}{g(\lambda_+-\lambda-)} \ .
\label{eq:coefficient2}
\end{align}
We assume that initially there is a localized region of excited atoms around the origin with width $l_s$. The initial amplitudes are taken to be 
\begin{align}
    \psi(\bx,0) & = 0\,\label{eq:init1} ,\\
    \sqrt{\rho_0}a(\bx,0) & = \left(\frac{1}{\pi l_s^2}\right)^{3/4}e^{-\vert\bx\vert^2/2l_s^2}. \label{eq:init2}
\end{align}
Taking the Fourier transform of ~(\ref{eq:init1}) and (\ref{eq:init2}) and using (\ref{eq:coefficient1}) and (\ref{eq:coefficient2}), we see that the components of $\bPsi(\bk,t)$ are given by
\begin{align}
    \hat{\psi}(\bk,t) &=\frac{g\rho_0 l_s^{3/2}}{2^{3/2}\pi^{5/2}}\frac{e^{-i\lambda_{+}(\bk)t}-e^{-i\lambda_{-}(\bk)t}}{\lambda_{+}(\bk)-\lambda_{-}(\bk)}e^{-l_s^2 \vert\bk\vert^2/2}\, ,\\
    \sqrt{\rho_0}\hat{a}(\bk,t)&=\frac{l_s^{3/2}}{2^{3/2}\pi^{5/2}}\frac{(\lambda_{+}(\bk)-\Omega)e^{-i\lambda_{-}(\bk)t}-(\lambda_{-}(\bk)-\Omega)e^{-i\lambda_{+}(\bk)t}}{\lambda_{+}(\bk)-\lambda_{-}(\bk)}e^{-l_s^2 \vert\bk\vert^2/2}.
\end{align}
Inverting the Fourier transforms, we find that 
\begin{align}
    \psi(\bx,t)&=\frac{g\sqrt{2}\rho_0 l_s^{3/2}}{\pi^{3/2}\vert\bx\vert}\int_0^{\infty}dk\,  k\sin(k\vert\bx\vert)\frac{e^{-i\lambda_{+}(k)t}-e^{-i\lambda_{-}(k)t}}{\lambda_{+}(k)-\lambda_{-}(k)}e^{-l_s^2 k^2/2}\, ,\\
    \sqrt{\rho_0}a(\bx,t)&=\frac{\sqrt{2}l_s^{3/2}}{\pi^{3/2}\vert\bx\vert}\int_0^{\infty}dk\, k\sin(k\vert\bx\vert)\frac{(\lambda_{+}(k)-\Omega)e^{-i\lambda_{-}(k)t}-(\lambda_{-}(k)-\Omega)e^{-i\lambda_{+}(k)t}}{\lambda_{+}(k)-\lambda_{-}(k)}e^{-l_s^2 k^2/2}.
\end{align}

\begin{figure}[t]
    \centering
    \includegraphics[width=4.in]{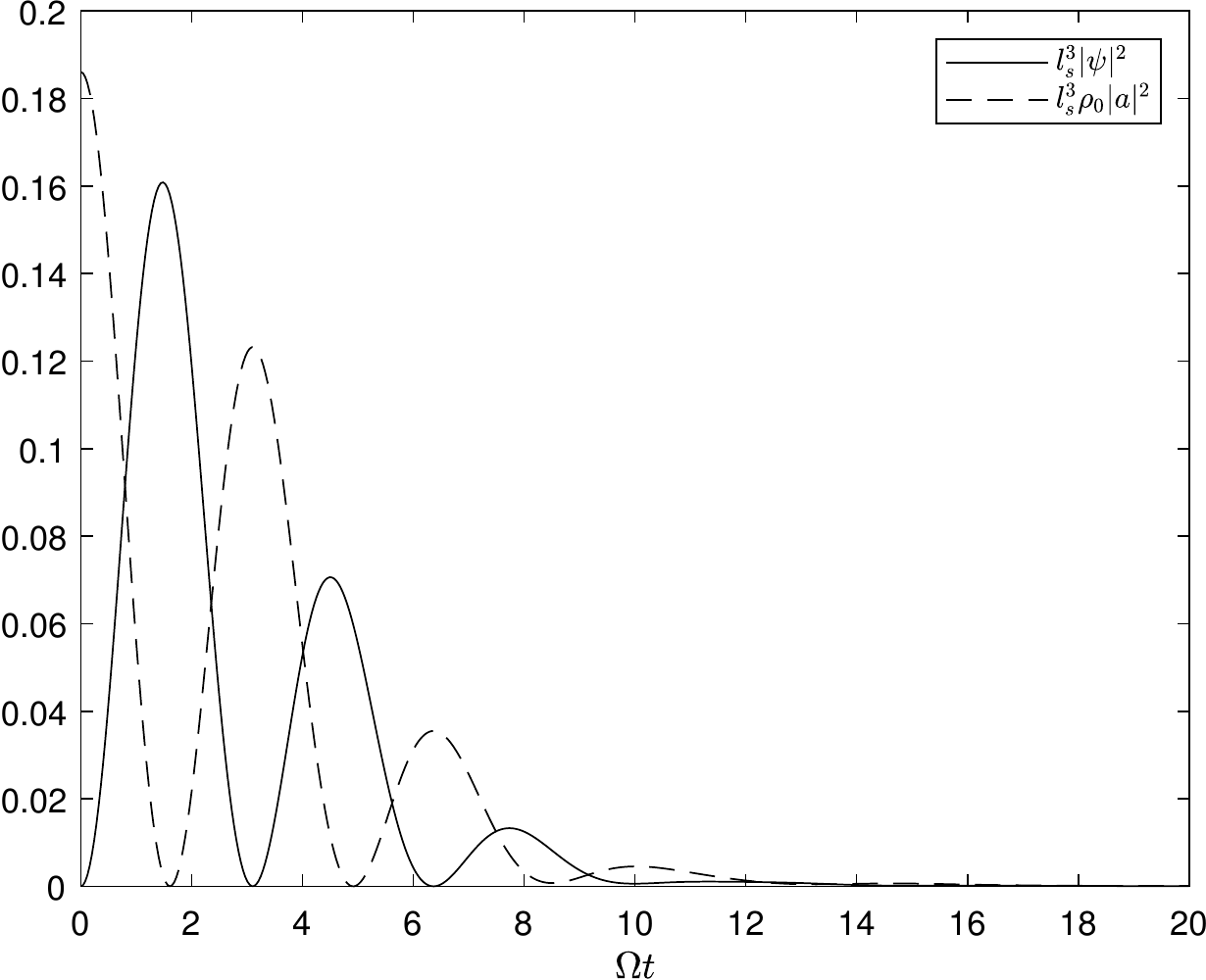}
    \caption{Time-dependence of atomic and field probability densities for the constant density problem with $|\bx|=l_s$}
    \label{fig:constantdensity}
\end{figure}

Figure~\ref{fig:constantdensity} illustrates the time-dependence of the probability densities $\vert\psi\vert^2$ and $\rho_0\vert a \vert^2$, where we have set the dimensionless quantities ${\Omega}/({\sqrt{\rho_0}\coup})={c}/({l_s\sqrt{\rho_0}\coup})=1$. We see that the probability densities are oscillatory and decay in time. 

\section{Energy Eigenstates}
\label{eigenstates}
\subsection{Radiative transport}
In this section we investigate the energy eigenstates of the Hamiltonian $H$ in a random medium. We consider the time-independent Schrodinger equation $H\vert\Psi\rangle=\hbar\omega\vert\Psi\rangle$, where $\vert\Psi\rangle$ is of the form (\ref{one_photon_state}) and $\hbar\omega$ is the energy. It follows that the amplitudes $a$ and $\psi$, which are independent of time, obey the equations
\begin{align}
\label{eq:psi}
c(-\Delta)^{1/2}\psi + \coup\rho(\bx) a = \omega\psi \ , \\
\coup\psi + \Omega a = \omega a \ .
\label{eq:a}
\end{align}
By eliminating $a$ from the above system, we immediately obtain the equation obeyed by $\psi$, which is given by
\begin{align} 
\label{eq:d2}
(-\Delta)^{1/2}\psi  + \frac{\coup^2\rho(\bx)}{c(\omega-\Omega)}\psi = k \psi \ ,
\end{align}
where $k=\omega/c$.

For the remainder of this work, we assume that the atomic density $\rho(\bx)$ is of the form
\begin{align}
\rho(\bx)=\rho_0(1+\eta(\bx)) \ ,
\end{align}
where $\rho_0$ is constant and $\eta(\bx)$ is a real-valued random field that accounts for statistical fluctuations in the density. We further assume that the correlations of $\eta$ are given by
\begin{align}
\langle\eta(\bx)\rangle &=0 \ , \\
\langle\eta(\bx)\eta(\by)\rangle &= C(\bx-\by) \ ,
\end{align}
where $C$ is the two-point correlation function and $\langle\cdots\rangle$ denotes statistical averaging. If $C$ depends only upon the quantity $\vert\bx-\by\vert$, the medium is said to be statistically homogeneous and isotropic. To make further progress, we consider the relative sizes of the important physical scales. The solution to (\ref{eq:d2}) oscillates on the scale of the wavelength $\lambda = 2\pi/k$. However, we are interested in the behavior of the solutions on the macroscopic scale $L\gg \lambda$.
We thus introduce a small parameter $\epsilon=\lambda/L$ and rescale the position $\bx$ by 
$\bx\to \bx/\epsilon$. In addition, we assume that the randomness is sufficiently weak so that the correlation function $C$ is $O(\epsilon)$. Thus (\ref{eq:d2}) becomes 
\begin{align}
\epsilon (-\Delta)^{1/2}\psi_{\epsilon} + k_0(1+\sqrt\epsilon\eta(\bx/\epsilon))\psi_{\epsilon} = k\psi_{\epsilon}  \ ,
\end{align}
where the $\epsilon$ dependence of $\psi$ is indicated explicitly  and
\begin{equation}
k_0 = \frac{\coup^2\rho_0}{c(\omega-\Omega)} \ .
\end{equation}
Note that we have also rescaled $\eta$ to be consistent with the $O(\epsilon)$ scaling of $C$.

We now introduce the Wigner transform of the amplitude $\psi$, which provides a phase-space representation of the correlation function of $\psi$. The Wigner transform $W_\epsilon(\bx,\bk)$ is defined as
\begin{align}
\label{wigner}
W_{\epsilon}(\bx,\bk)=\int\frac{d^3 x'}{(2\pi)^3}e^{-i\bk\cdot\bx'}\psi_{\epsilon}(\bx-\epsilon\bx'/2)\psi_{\epsilon}^*(\bx+\epsilon\bx'/2) \ .
\end{align}
The Wigner transform has several important properties. It is real-valued and related to the probability density $|\psi_\epsilon|^2$ by
\begin{align}
\vert\psi_\epsilon(\bx)\vert^2 = \int d^3 k W_\epsilon(\bx,\bk) \ .
\end{align}

Next we derive a useful relation governing the Wigner transform. Let $\Phi_{\epsilon}(\bx_1,\bx_2)=\psi_{ \epsilon}(\bx_1)\psi_{ \epsilon}^*(\bx_2)$. Since $\eta$ is real-valued, it follows that $\Phi_{\epsilon}(\bx_1,\bx_2)$ satisfies the pair of equations
\begin{align} \label{eq:d4}
\epsilon (-\Delta_{\bx_1})^{1/2}\Phi_{\epsilon}(\bx_1,\bx_2) -k\Phi_{\epsilon}(\bx_1,\bx_2) + k_0(1+\sqrt{\epsilon}\eta(\bx_1/\epsilon))\Phi_{\epsilon}(\bx_1,\bx_2) =0 \ ,
\end{align}
\begin{align} \label{eq:d5}
\epsilon (-\Delta_{\bx_2})^{1/2}\Phi_{\epsilon}(\bx_1,\bx_2) -k\Phi_{\epsilon}(\bx_1,\bx_2) + k_0(1+\sqrt{\epsilon}\eta(\bx_2/\epsilon))\Phi_{\epsilon}(\bx_1,\bx_2) =0 \ .
\end{align}
Subtracting (\ref{eq:d4}) from (\ref{eq:d5}) yields
\begin{align}
\epsilon \left[(-\Delta_{\bx_1})^{1/2}-(-\Delta_{\bx_2})^{1/2}\right]\Phi_{\epsilon}(\bx_1,\bx_2)+\sqrt{\epsilon}k_0\left[\eta(\bx_1/\epsilon)-\eta(\bx_2/\epsilon)\right]\Phi_{\epsilon}(\bx_1,\bx_2)=0.
\end{align}
We now perform the change of variables 
\begin{align}
    \bx_1&=\bx-\epsilon\bx'/2 \ , \\
    \bx_2&=\bx+\epsilon\bx'/2 \ ,
\end{align}
and Fourier transform the result with respect to $\bx'$, thus arriving at 
\begin{align} 
\label{eq:d6}
\int\frac{d^3 q}{(2\pi)^3}e^{i\bq\cdot\bx}\left[\vert -\bk+\epsilon\bq/2\vert-\vert\bk+\epsilon\bq/2\vert\right]\tilde{W}_{\epsilon}(\bq,\bk)+ \sqrt{\epsilon}LW_{\epsilon}(\bx,\bk)=0 \ ,
\end{align}
where the Fourier transform of the Wigner transform is defined by
\begin{align}
\tilde{W}_{\epsilon}(\bq,\bk)=\int d^3x e^{-i\bq\cdot\bx}\,W_{\epsilon}(\bx,\bk)
\end{align}
and
\begin{align}
   LW_{\epsilon}(\bx,\bk)=k_0\int\frac{d^3 q}{(2\pi)^3}e^{i\bq\cdot\bx/\epsilon}\tilde{\eta}(\bq)\left[W_{\epsilon}(\bx,\bk+\bq/2)-W_{\epsilon}(\bx,\bk-\bq/2)\right].
\end{align}
The details of the calculation are given in Appendix B. 

We now consider the behavior of $W_\epsilon$ in the high-frequency limit $\epsilon\to 0$, which allows for the separation of microscopic and macroscopic scales. To this end we introduce a multiscale expansion for $W_\epsilon$ of the form

\begin{align} 
\label{eq:d7}
    W_{\epsilon}(\bx,\bk)= W_0(\bx,\bk)+\sqrt{\epsilon}W_1(\bx,\bX,\bk)+\epsilon W_2(\bx,\bX,\bk)+\cdots \ ,
\end{align}
where $\bX=\bx/\epsilon$ is a fast variable and $W_0$ is taken to be deterministic. We treat $\bx$ and $\bX$ as independent variables and make the replacement
\begin{align} 
\nabla_{\bx}\to\nabla_{\bx}+\frac{1}{\epsilon}\nabla_{\bX} \ .
\label{derivatives}
\end{align}
Eq.~(\ref{eq:d6}) thus becomes
\begin{align} 
\label{eq:d8}
\int\frac{d^3 q}{(2\pi)^3}\frac{d^3 Q}{(2\pi)^3}e^{i\bq\cdot\bx+i\bQ\cdot\bX}\left[\vert -\bk+\epsilon\bq/2+\bQ/2\vert-\vert\bk+\epsilon\bq/2+\bQ/2\vert\right]\tilde{W}_{\epsilon}(\bq,\bQ,\bk)\\
+ \sqrt{\epsilon}LW_{\epsilon}(\bx,\bX,\bk)=0 \ ,
\end{align}
where
\begin{align}
\tilde{W}_{\epsilon}(\bq,\bQ,\bk) =\int d^3 xd^3 Xe^{-i\bq\cdot\bx-i\bQ\cdot\bX}\,W_{\epsilon}(\bx,\bX,\bk).
\end{align}
Inserting (\ref{eq:d7}) into (\ref{eq:d8}) and equating terms of the same order in $\epsilon$, we find that at $O(\sqrt{\epsilon})$ 
\begin{align} \label{eq:d9}
\int\frac{d^3 Q}{(2\pi)^3}e^{i\bQ\cdot\bX}\left[\vert-\bk+\bQ/2\vert-\vert\bk+\bQ/2\vert\right]\tilde{W}_{1}(\bx,\bQ,\bk)+\sqrt{\epsilon}LW_{0}(\bx,\bk)=0.
\end{align}
Eq.~(\ref{eq:d9}) can be solved by Fourier transforms with the result
\begin{align} 
\label{eq:d10}
\tilde{W}_1(\bx,\bQ,\bk)=k_0\tilde{\eta}(\bQ)\frac{[W_0(\bx,\bk+\bQ/2)-W_0(\bx,\bk-\bQ/2)]}{\vert\bk+\bQ/2\vert-\vert -\bk+\bQ/2\vert-i\theta} \ ,
\end{align}
where $\theta\to 0$ is a positive regularizing parameter. At order $O(\epsilon)$ we find that
\begin{align} 
\label{eq:d11}
&\int\frac{d^3 q}{(2\pi)^3}\frac{d^3 Q}{(2\pi)^3}e^{i\bq\cdot\bx + i\bQ\cdot\bX}\left[\vert -\bk+\bQ/2\vert-\vert\bk+\bQ/2\vert\right]\tilde{W_2}(\bq,\bQ,\bk)-\int\frac{d^3 q}{(2\pi)^3}e^{i\bq\cdot\bx}\hat\bk\cdot\bq\,\tilde{W_0}(\bq,\bk) 
\nonumber \\
+& k_0\int\,\frac{d^3 q}{(2\pi)^3}e^{i\bq\cdot\bX}\tilde{\eta}(\bq)\left[W_1(\bx,\bX,\bk+\bq/2)-W_1(\bx,\bX,\bk-\bq/2)\right]=0 \ .
\end{align}

Next we average (\ref{eq:d11}) over realizations of the random medium. To do so, we impose the condition $\langle\left[\vert -\bk+\bQ/2\vert-\vert\bk+\bQ/2\vert\right]\tilde{W_2}(\bq,\bQ,\bk)\rangle=0$, which closes the hierarchy relating the terms in the multiscale expansion, and corresponds to the assumption that $W_2$ is statistically stationary in the fast variable $\bX$. Eq.~(\ref{eq:d11}) thus becomes
\begin{align} 
\label{eq:d12}
\hat\bk\cdot\nabla_{\bx}\,W_0(\bx,\bk)+k_0\int\frac{d^3 q}{(2\pi)^3}e^{i\bq\cdot\bX}\langle\tilde{\eta}(\bq)\left[W_1(\bx,\bX,\bk+\bq/2)-W_1(\bx,\bX,\bk-\bq/2)\right]\rangle=0.
\end{align}
After substituting ~(\ref{eq:d10}) into ~(\ref{eq:d12}) and using the identity 
\begin{align}
 \langle\tilde{\eta}(\bp)\tilde{\eta}(\bq)\rangle=(2\pi)^3\delta(\bp+\bq)\tilde{C}(\bp)
\end{align}
we find, as shown in Appendix C, that $W_0$ satisfies
\begin{align} 
\label{eq:d13}
\nonumber
\hat\bk\cdot\nabla_{\bx}W_0(\bx,\bk)&+k_0^2\int\frac{d^3 q}{(2\pi)^2}\,\tilde{C}(\bq-\bk)\delta(\vert\bq\vert-\vert\bk\vert)W_0(\bx,\bk)\\
    &=k_0^2\int\frac{d^3 q}{(2\pi)^2}\,\tilde{C}(\bq-\bk)\delta(\vert\bq\vert-\vert\bk\vert)W_0(\bx,\bq) \ .
\end{align}
Here we define the scattering coefficient $\mu_s$ and phase function $A$ as
\begin{align}
    \mu_s &= \frac{k_0^2\vert \bk\vert^2}{4\pi^2}\int d \hbk'\,\tilde{C}\left(\vert\bk\vert(\hbk-\hbk')\right) \ , \\
    A(\hbk,\hbk')&=\frac{k_0^2\vert \bk\vert^2}{\mu_s}\tilde{C}\left(\vert\bk\vert(\hbk-\hbk')\right)
\end{align}
Making use of these definitions, (\ref{eq:d13}) becomes
\begin{align}
\label{RTE_time_independent}
\hbk\cdot\nabla_{\bx}W_0(\bx,\bk)+\mu_s W_0(\bx,\bk)=\mu_s LW_0(\bx,\bk) \ ,
\end{align}
where the operator $L$ is defined by
\begin{align}
    LW_0(\bx,\bk)=\int\,d \hbk' A(\hbk,\hbk')W_0(\bx,\bk').
\end{align}

Eq.~(\ref{RTE_time_independent}), which has the form of a time-independent radiative transport equation, is the main result of this section. We note that $\mu_s$ and $A$ are defined in terms of the correlations of the medium. Since the density fluctuations $\eta$ are statistically homogeneous and isotropic, $\tilde{C}$ depends only on the quantity $\vert\bk-\bk'\vert$, and hence the phase function $A$ depends only on $\hbk\cdot\hbk'$ and $\vert\bk\vert$. Similarly, $\mu_s$ only depends on the magnitude $\vert\bk\vert$.

In the case of white noise-disorder, where $C=C_0\delta(\bx)$ with constant $C_0$, the scattering coefficient and phase function are given by
\begin{align}
    \mu_s &= 4\pi C_0 k_0^2 \vert\bk\vert^2 \ , \\
    A(\hbk,\hbk')&=\frac{1}{4\pi} \ ,
\end{align}
which corresponds to isotropic scattering.

\subsection{Diffusion Approximation}

We now consider the diffusion limit of the radiative transport equation developed in the previous section. The diffusion approximation for a radiative transport equation of the form
\begin{align} \label{eq:e1}
\hbk\cdot\nabla_{\bx}W_0(\bx,\bk)+\mu_sW_0(\bx,\bk)=\mu_sLW_0(\bx,\bk) 
\end{align}
is obtained by expanding $W_0$ in spherical harmonics~\cite{Carminati_2020}. To lowest order, it can be seen that
\begin{align}
    W_0(\bx,\bk)=\frac{1}{4\pi}\left(u(\bx,\vert\bk\vert)-\ell^*\hbk\cdot\nabla u(\bx,\vert\bk\vert)\right) \ ,
\end{align}
where the first angular moment $u(\bx,\vert\bk\vert)$ is defined by
\begin{align}
u(\bx,\vert\bk\vert) = \int d\hbk W_0(\bx,\bk) \ ,
\end{align}
and the transport mean free path $\ell^*$ is given by
\begin{align} \label{eq:e2}
    \ell^*=\frac{1}{\mu_s(1-g)}, \quad g = \int d\hbk' \hbk\cdot\hbk' A(\hbk,\hbk') \ .
\end{align}
The anisotropy $g$ takes values between $-1$ and $1$ and vanishes for isotropic scattering. The quantity $u$ satisfies the diffusion equation 
\begin{align}
\Delta u &= 0 \quad {\rm in} \quad \Omega  \ , \\
u &= g \quad {\rm on} \quad \partial\Omega \ ,
\end{align}
where we have prescribed Dirichlet boundary conditions on a bounded domain $\Omega$ and $g$ generally depends upon $k$. Since $\vert\psi\vert^2$ is given by 
\begin{align}
    \vert\psi(\bx)\vert^2 = \int\dd^3 k\, W_0(\bx,\bk) = \int_0^{\infty}\dd k\,k^2 u(\bx,k) \ ,
\end{align}
it follows that $\vert\psi\vert^2$ obeys
\begin{align}
\Delta \vert\psi\vert^2 &= 0 \quad \text{in} \quad \Omega \ , \\
\vert\psi\vert^2 &=\int_0^{\infty} dk k^2 g(\bx,k) \quad \text{on}\quad \partial\Omega \ ,
\end{align}
where the $k$ dependence of $g$ has been made explicit.

\section{Collective Spontaneous Emission}
\label{collective}
\subsection{Kinetic equations}

In this section we study the time evolution of the atomic and field amplitudes in a random medium. Our starting point is (\ref{eq:b4}) and (\ref{eq:b5}) (with $\rho$ cancelled):
\begin{align} 
\label{eq:psitime}
i\partial_t\psi &= c(-\Delta)^{1/2}\psi + \coup\rho(\bx) a\ , \\
i\partial_t a &= \coup\psi + \Omega a \ .
\label{eq:atime}
\end{align}
A similar system of pseudodifferential equations with a random potential has been considered in ~\cite{Bal_2005}. If we define the vector quantity $\bu(\bx,t)=\begin{bmatrix} \psi(\bx,t) , a(\bx,t)\sqrt{\rho_0} \end{bmatrix}^{T}$, then $\bu$ satisfies the equation
\begin{align} \label{eq:system}
i\partial_t\bu = A(\bx)\bu+\coup\sqrt{\rho_0}\eta(\bx)K\bu\ ,
\end{align}
where
\begin{align} 
A(\bx) &= \begin{bmatrix}c(-\Delta_\bx)^{1/2}& \coup\sqrt{\rho_0}\\ \coup\sqrt{\rho_0} & \Omega \end{bmatrix} \ , \\
K &= \begin{bmatrix} 0 & 1\\ 0 & 0 \end{bmatrix}.
\end{align}
This definition of $\bu$ has the advantage that its two components have the same dimensions and that the matrix $A(\bx)$ is symmetric. We perform the same rescaling of the variables $\bx$ and $\eta$ as previously, and we also rescale the time $t$ as $t\to t/\epsilon$. Thus (\ref{eq:system}) becomes
\begin{align}
\epsilon i\partial_t\bu_{\epsilon} = A_{\epsilon}(\bx)\bu_{\epsilon}+\sqrt{\epsilon}\coup\sqrt{\rho_0}\eta(\bx/\epsilon)K\bu_{\epsilon}\ ,
\end{align}
where
\begin{align}
A_{\epsilon}(\bx)=\begin{bmatrix}\epsilon c(-\Delta_\bx)^{1/2}& \coup\sqrt{\rho_0}\\ \coup\sqrt{\rho_0} & \Omega \end{bmatrix} \ .
\end{align}

We now consider the Wigner transform of $\bu$, which is  matrix-valued and defined
by
\begin{align}
    W_{\epsilon}(\bx,\bk,t)=\int\frac{d^3 x'}{(2\pi)^3}e^{-i\bk\cdot\bx'}\bu_{\epsilon}(\bx-\epsilon\bx'/2,t)\bu_{\epsilon}^{\dagger}(\bx+\epsilon\bx'/2,t) \ .
\end{align}
The probability densities $\vert\psi_{\epsilon}(\bx,t)\vert^2$ and $\vert a_{\epsilon}(\bx,t)\vert^2$ are related to the Wigner transform by
\begin{align}
\label{prob_density_psi}
\vert\psi_{\epsilon}(\bx,t)\vert^2 &=\int d^3 k (W_{\epsilon})_{11}(\bx,\bk,t) \ , \\
\rho_0\vert a_{\epsilon}(\bx,t)\vert^2 &= \int d^3 k (W_{\epsilon})_{22}(\bx,\bk,t) \ .
\label{prob_density_a}
\end{align}
If we define $\Phi_{\epsilon}(\bx_1,\bx_2,t)=\bu_{\epsilon}(\bx_1,t)\bu_{\epsilon}^*(\bx_2,t)$, then $\Phi_{\epsilon}$ satisfies the equation
\begin{align}
\nonumber
\epsilon i\partial_t\Phi_{\epsilon}&= A_{\epsilon}(\bx_1)\Phi_{\epsilon}+\sqrt{\epsilon}\coup\sqrt{\rho_0}\eta(\bx_1/\epsilon)K\Phi_{\epsilon}\\
    &-\Phi_{\epsilon}A_{\epsilon}(\bx_2)+\sqrt{\epsilon}\coup\sqrt{\rho_0}\eta(\bx_2/\epsilon)\Phi_{\epsilon}K^{T}.
\end{align}
Next we perform the change of variables 
\begin{align}
    \bx_1&=\bx-\epsilon\bx'/2 \ , \\
    \bx_2&=\bx+\epsilon\bx'/2 \ ,
\end{align}
and Fourier transform the result with respect to $\bx'$. We thus obtain
\begin{align} \label{eq:f1}
\nonumber\epsilon i\partial_t W_{\epsilon}(\bx,\bk,t) =&\int\frac{d^3 q}{(2\pi)^3}e^{i\bq\cdot\bx}\left[\tilde{A}_{\epsilon}(\bk/\epsilon-\bq/2)\tilde{W}_{\epsilon}(\bq,\bk,t)-\tilde{W}_{\epsilon}(\bq,\bk,t)\tilde{A}_{\epsilon}(\bk/\epsilon+\bq/2)\right]\\
+&\sqrt{\epsilon}\coup\sqrt{\rho_0}\int\frac{d^3 q}{(2\pi)^3}e^{i\bq\cdot\bx/\epsilon}\tilde{\eta}(\bq)\left[KW_{\epsilon}(\bx,\bk+\bq/2,t)-W_{\epsilon}(\bx,\bk-\bq/2,t)K^{T}\right] \ ,
\end{align}
where
\begin{equation}
\tilde{A}_{\epsilon}(\bk) = \begin{bmatrix}\epsilon c\vert\bk\vert& \coup\sqrt{\rho_0}\\ \coup\sqrt{\rho_0} & \Omega \end{bmatrix} \ .
\end{equation}
The details of this calculation are given in Appendix D. 

Once again we consider the behavior of $W_\epsilon$ in the high-frequency limit $\epsilon\to 0$. To this end we introduce a multiscale expansion for $W_\epsilon$ of the form
\begin{align} 
\label{eq:f2}
W_{\epsilon}(\bx,\bk,t)= W_0(\bx,\bk,t)+\sqrt{\epsilon}W_1(\bx,\bX,\bk,t)+\epsilon W_2(\bx,\bX,\bk,t)+\cdots \ ,
\end{align}
where $\bX=\bx/\epsilon$ is a fast variable, and $W_0$ is taken to be deterministic and independent of $\bX$. We treat $\bx$ and $\bX$ as independent variables and transform the derivative $\nabla_{\bx}$ according to (\ref{derivatives}).
Eq.~(\ref{eq:f1}) thus becomes
\begin{align} 
\label{eq:f3}
    &\nonumber\epsilon i\partial_t W_{\epsilon}(\bx,\bX,\bk,t)=\int\frac{d^3 q}{(2\pi)^3}\frac{d^3 Q}{(2\pi)^3}e^{i\bq\cdot\bx+i\bQ\cdot\bX}\left[\tilde{A}_{\epsilon}(\bk/\epsilon-\bq/2-\bQ/2\epsilon)\tilde{W}_{\epsilon}(\bq,\bQ,\bk,t)\right.\\
    &\nonumber\left.-\tilde{W}_{\epsilon}(\bq,\bQ,\bk,t)\tilde{A}_{\epsilon}(\bk/\epsilon+\bq/2+\bQ/2\epsilon)\right]
    +\sqrt{\epsilon}\coup\sqrt{\rho_0}\int\frac{d^3 q}{(2\pi)^3}e^{i\bq\cdot\bX}\left[KW_{\epsilon}(\bx,\bX,\bk+\bq/2,t)\right.\\
    &\left.-W_{\epsilon}(\bx,\bX,\bk-\bq/2,t)K^{T}\right].
\end{align}
Inserting (\ref{eq:f2}) into (\ref{eq:f3}) and equating terms of the same order in $\epsilon$, we find that at $O(1)$
\begin{align} \label{eq:f4}
    \tilde{A}_{\epsilon}(\bk/\epsilon)W_{0}(\bx,\bk,t)-W_{0}(\bx,\bk,t)\tilde{A}_{\epsilon}(\bk/\epsilon)=0.
\end{align}
Since $\tilde{A}_{\epsilon}(\bk/\epsilon)$ is symmetric it can be diagonalized. Its eigenvalues are given by
\begin{align} 
\label{eigenval}
    \lambda_{\pm}(\bk) &=\frac{(c\vert\bk\vert+\Omega)\pm\sqrt{(c\vert\bk\vert-\Omega)^2+4\coup^2\rho_0}}{2} \ .
   \end{align}
The corresponding eigenvectors are real and are of the form
\begin{align} 
\label{eigenvec}
    \bb_{\pm}(\bk) &=\frac{1}{\sqrt{(\lambda_{\pm}-\Omega)^2+\coup^2\rho_0}}\begin{bmatrix}
                \lambda_{\pm}-\Omega\\
                \coup\sqrt{\rho_0}
                \end{bmatrix}.
 \end{align}
It follows from (\ref{eq:f4}) that $W_0$ is also diagonal in the basis $\{\bb_+(\bk),\bb_-(\bk)\}$ and can be expressed as
\begin{align} 
\label{eq:f5}
    W_0(\bx,\bk,t)=\modei(\bx,\bk,t)\bb_+(\bk)\bb_+^{T}(\bk)+\modeii(\bx,\bk,t)\bb_-(\bk)\bb_-^{T}(\bk) \ ,
\end{align}
where $a_\pm$ are suitable coefficients.

At order $O(\sqrt{\epsilon})$ we obtain
\begin{align}
\label{eq:f6}
    &\tilde{A}_{\epsilon}((\bk-\bQ/2)/\epsilon)\tilde{W}_{1}(\bx,\bQ,\bk,t)-\tilde{W}_{1}(\bx,\bQ,\bk,t)\tilde{A}_{\epsilon}((\bk+\bQ/2)/\epsilon) \\
&=\coup\sqrt{\rho_0}\tilde{\eta}(\bq)\left[W_{0}(\bx,\bk-\bQ/2,t)K^{T}-KW_{0}(\bx,\bk+\bQ/2,t)\right] \ .
\end{align}
We can then decompose $\tilde{W}_1$ as
\begin{align}\label{eq:f7}
\tilde{W}_1(\bx,\bQ,\bk,t) = \sum_{m,n}w_{mn}(\bx,\bQ,\bk,t)\bb_m(\bk-\bQ/2)\bb_n^T(\bk+\bQ/2) \ ,
\end{align}
for suitable coefficients $w_{mn}$.
Multiplying (\ref{eq:f6}) on the left by $\bb_m^{T}(\bk-\bQ/2)$, on the right by $\bb_n(\bk+\bQ/2)$, and using the facts 
\begin{align}
\bb_m^{T}(\bq)K\bb_n(\bp) &=\frac{\coup\sqrt{\rho_0}(\lambda_m(\bq)-\Omega)}{\sqrt{(\lambda_m(\bq)-\Omega)^2+\coup^2\rho_0}\sqrt{(\lambda_n(\bp)-\Omega)^2+\coup^2\rho_0}} \ , \\
\bb_m^{T}(\bq)K^{T}\bb_n(\bp) &=\frac{\coup\sqrt{\rho_0}(\lambda_n(\bp)-\Omega)}{\sqrt{(\lambda_m(\bq)-\Omega)^2+\coup^2\rho_0}\sqrt{(\lambda_n(\bp)-\Omega)^2+\coup^2\rho_0}} \ ,
\end{align}
we find that
\begin{align}
\label{eq:f8}
    &\nonumber w_{mn}(\bx,\bQ,\bk,t)\\
    =& \frac{\coup^2\rho_0\tilde{\eta}(\bQ)((\lambda_n(\bk+\bQ/2)-\Omega)\modem(\bx,\bk-\bQ/2,t)-(\lambda_m(\bk-\bQ/2)-\Omega)\moden(\bx,\bk+\bQ/2,t)}{\sqrt{(\lambda_m(\bk-\bQ/2)-\Omega)^2+\coup^2\rho_0}\sqrt{(\lambda_n(\bk+\bQ/2)-\Omega)^2+\coup^2\rho_0}(\lambda_{m}(\bk-\bQ/2)-\lambda_{n}(\bk+\bQ/2)+i\theta)} \ ,
\end{align}
where $\theta\to 0$ is a positive regularizing parameter. 
At order $O(\epsilon)$ we obtain
\begin{align} 
\label{eq:f9}
\nonumber 
i\partial_t W_{0}(\bx,\bk,t) &=LW_2(\bx,\bX,\bk,t)+M(\bx,\bk)W_{0}(\bx,\bk,t)\\
&+\coup\sqrt{\rho_0}\int\frac{d^3 q}{(2\pi)^3}e^{i\bq\cdot\bX}\tilde{\eta}(\bq)\left[KW_{1}(\bx,\bX,\bk+\bq/2,t)-W_{1}(\bx,\bX,\bk-\bq/2,t)K^{T}\right] \ ,
\end{align}
where 
\begin{align}
    LW_2(\bx,\bX,\bk,t) &= \int\frac{d^3 q}{(2\pi)^3}\frac{d^3 Q}{(2\pi)^3}e^{i\bq\cdot\bx+i\bQ\cdot\bX}\left[\tilde{A}_{\epsilon}(\bk/\epsilon-\bQ/2\epsilon)\tilde{W}_{2}(\bq,\bQ,\bk,t)\right.\\
    \nonumber&\left.-\tilde{W}_{2}(\bq,\bQ,\bk,t)\tilde{A}_{\epsilon}(\bk/\epsilon+\bQ/2\epsilon)\right]\ ,\\
    M(\bx,\bk)&=\begin{bmatrix}ic\ \hbk\cdot\nabla_{\bx} & 0 \\
        0 & 0 \end{bmatrix}.
\end{align}

In order to obtain the equation satisfied by $\modei$ ($\modeii$), we multiply (\ref{eq:f9}) on the left by $\bb_{+}^{T}(\bk)$ ($\bb_{-}^{T}(\bk)$) and on the right by $\bb_+(\bk)$ ($\bb_-(\bk)$) and take the average. Moreover, we assume that $\langle \bb_{\pm}^{T} LW_2\bb_{\pm}\rangle=0$, which closes the hierarchy of equations and corresponds to the assumption that $W_2$ is statistically stationary in the fast variable $\bX$. This leads to the kinetic equations
\begin{align}
\label{RTE_modes}
\nonumber
\frac{1}{c}\partial_t a_{\pm}(\bx,\bk,t)
&+f_{\pm}(\bk)\hbk\cdot\nabla_{\bx}a_{\pm}(\bx,\bk,t)+\mu_{\pm}(\bk)a_{\pm}(\bx,\bk,t)\\
&=\mu_{\pm}(\bk)\int d\hbk'A(\bk,\bk')a_{\pm}(\bx,\bk',t) \ ,
\end{align}
which is the main result of this paper.
Here the scattering coefficients $\mu_{\pm}$, the phase function $A$ and transport coefficients $f_{\pm}$ are defined by
\begin{align}
\label{}
    \mu_{\pm}(\bk) &=\frac{4\pi (\coup^2\rho_0)^2\vert\lambda_{\pm}(\bk)-\Omega\vert}{c^2((\lambda_{\pm}(\bk)-\Omega)^2+\coup^2\rho_0)^2}\sqrt{(c\vert\bk\vert-\Omega)^2+4\coup^2\rho_0}\vert\bk\vert^2\int\frac{d\hbk'}{(2\pi)^3}\,\tilde{C}(\vert\bk\vert(\hbk-\hbk')) \ , \\
    A(\bk,\bk') &=\frac{\tilde{C}(\vert\bk\vert(\hbk-\hbk'))}{\displaystyle\int d\hbk'\tilde{C}(\vert\bk\vert(\hbk-\hbk'))} \ , \\
    f_{\pm}(\bk)&=\frac{(\lambda_{\pm}(\bk)-\Omega)^2}{(\lambda_{\pm}(\bk)-\Omega)^2+\coup^2\rho_0} \ .
\end{align}
The details of this calculation are given in Appendix E. 

Suppose that $\psi$ and $a$ have time dependences
\begin{align}
\psi(\bx,t)=e^{-i\omega t}\psi_0(\bx) \ , \quad  a(\bx,t) =e^{-i\omega t}a_0(\bx) \ ,
\end{align}
which correspond to eigenstates of the Hamiltonian with energy $\hbar\omega$. 
Then using (\ref{RTE_modes}), it can be seen that the Wigner transforms of $\psi_0$ and $a_0$ satisfy the radiative transport equation (\ref{RTE_time_independent}). That is, the results for the time-independent problem are consistent with those of the time-dependent problem.

The Wigner transform $W_0$ can be obtained from the solution to the RTE (\ref{RTE_modes}) by making use of (\ref{eq:f5}). It follows from (\ref{prob_density_psi}) and (\ref{prob_density_a}) that  the average probability densities $\langle\vert\psi\vert^2\rangle$ and $\langle\vert a\vert^2\rangle$ are given by
\begin{align}
\label{eq:f11}
\nonumber
\langle\vert\psi(\bx,t)\vert^2\rangle &=\int d^3 k (W_0)_{11}(\bx,\bk,t) \\
& =\int d^3 k\left[\frac{\modei(\bx,\bk,t)(\lambda_+(\bk)-\Omega)^2}{(\lambda_+(\bk)-\Omega)^2+\coup^2\rho_0 }+\frac{\modeii(\bx,\bk,t)(\lambda_-(\bk)-\Omega)^2}{(\lambda_-(\bk)-\Omega)^2+\coup^2\rho_0}\right] \ , \\ \nonumber \\
\nonumber
\rho_0\langle\vert a(\bx,t)\vert^2\rangle &=\int d^3 k (W_0)_{22}(\bx,\bk,t) \\
& =\coup^2\rho_0\int d^3 k\left[\frac{\modei(\bx,\bk,t)}{(\lambda_+(\bk)-\Omega)^2+\coup^2\rho_0 }+\frac{\modeii(\bx,\bk,t)}{(\lambda_-(\bk)-\Omega)^2+\coup^2\rho_0}\right] \ .
\label{eq:f111}
\end{align}

\subsection{Diffusion Approximation}
We now consider the diffusion approximation to the kinetic equation (\ref{RTE_modes}). The diffusion approximation for a kinetic equation of the form
\begin{align} 
\label{eq:g1}
\frac{1}{c}\partial_t I(\bx,\bk,t) + \hbk\cdot\nabla_{\bx}I(\bx,\bk,t)+\mu_sI(\bx,\bk,t)=\mu_s LI(\bx,\bk,t)
\end{align}
is obtained by expanding $I$ in spherical harmonics~\cite{Carminati_2020}. To lowest order, it can be seen that
\begin{align}
    I(\bx,\bk,t)=\frac{1}{4\pi}\left(u(\bx,\vert\bk\vert,t)-\ell^*\hbk\cdot\nabla u(\bx,\vert\bk\vert,t)\right) \ ,
\end{align}
where $u(\bx,\vert\bk\vert,t)$ is defined by
\begin{align}
u(\bx,\vert\bk\vert,t) = \int d\hbk I(\bx,\bk,t) \ ,
\end{align}
and $\ell^*$, which depends on $\vert\bk\vert$, is defined by (\ref{eq:e2}).
We then find that $u$ satisfies the diffusion equation
\begin{align}
\label{diff_eq}
\partial_t u = D\Delta u \ ,
\end{align}
where the diffusion coefficient $D$ is given by
\begin{align}
D=\frac{1}{3}c\ell^* \ .
\end{align}
The solution to (\ref{diff_eq}) for an infinite medium is given by
\begin{align}
\label{soln}
u(\bx,t) = \frac{1}{(4\pi D t)^{3/2}}\int d^3x'  \exp\left[-\frac{|\bx-\bx'|^2}{4D t}\right]u(\bx',0) \ .
\end{align}
We note that the diffusion approximation is accurate at large distances and long times.

It follows from the above that the first angular moments of $a_\pm$, which are defined by
\begin{align}
    u_{\pm}(\bx,\vert\bk\vert,t)&=\int d\hbk a_{\pm}(\bx,\bk,t) \ ,
\end{align}
satisfy diffusion equations of the form
\begin{align} \label{eq:g2}
\partial_t u_{\pm} = D_{\pm}(k)\Delta u_{\pm} \ .
\end{align}
Here the diffusion coefficients are given by
\begin{align}
D_{\pm}(k) = \frac{cf_{\pm}(k)^2}{3(1-g)\mu_{\pm}(k)} \ .
\end{align}

\begin{figure}[t]
\begin{center}
\includegraphics[width=4.in]{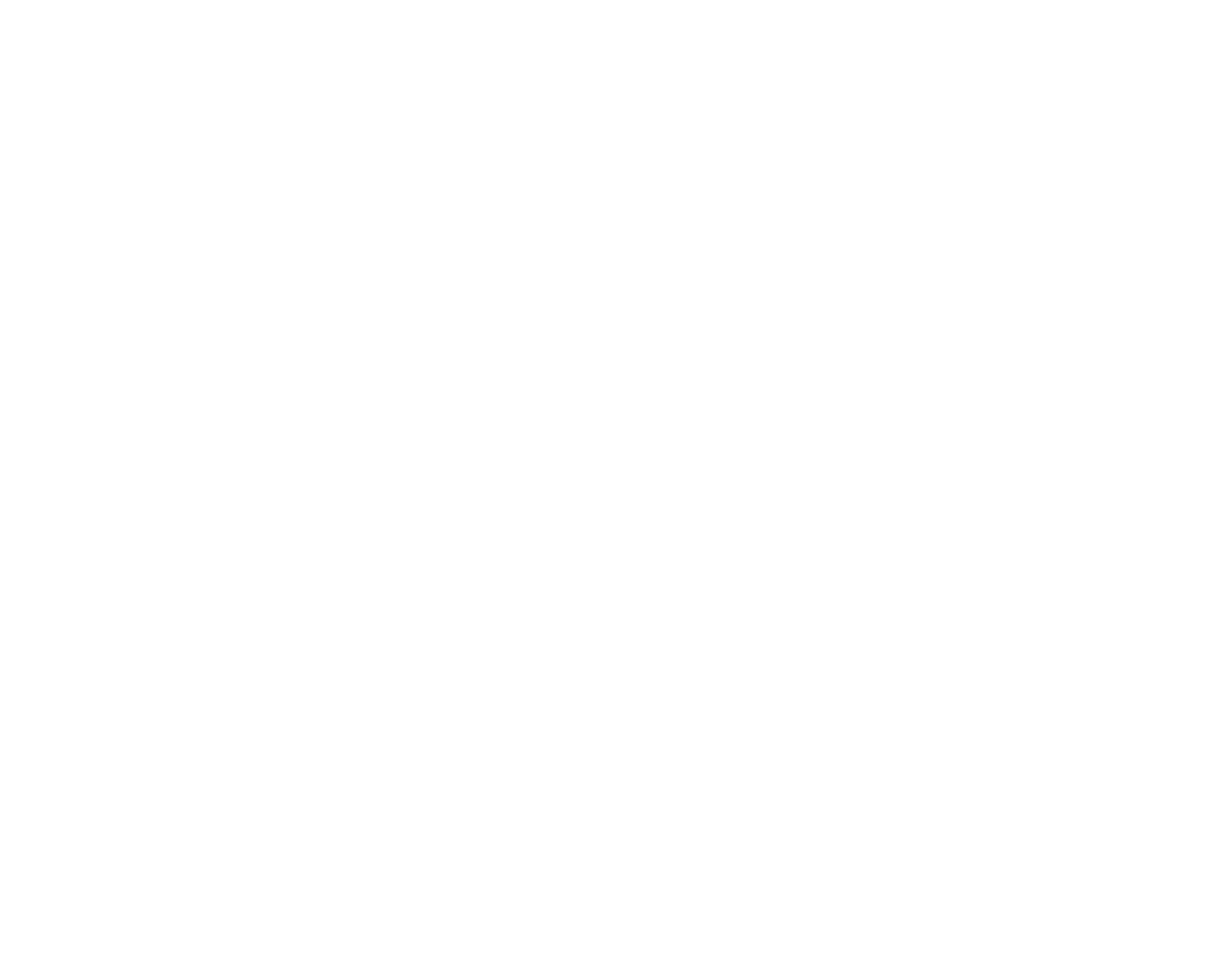}
\end{center}
\caption{\label{fig:multiplepositions}Time dependence of atomic probability density in a random medium for several distances $|\bx|$.}
\end{figure}

In order to compute $\langle\vert\psi(\bx,t)\vert^2\rangle$ and $\langle\vert a(\bx,t)\vert^2\rangle$ from (\ref{eq:f11}) and (\ref{eq:f111}), we must specify the initial conditions $\psi(\bx,0)$ and $a(\bx,0)$, which in turn imply initial conditions on $a_\pm$ of the form
\begin{align} 
\label{eq:f12a}
\nonumber
\modei(\bx,\bk,0)&=\frac{(W_0)_{11}(\bx,\bk,0)\left[(\lambda_+(\bk)-\Omega)^2+\coup^2\rho_0\right]}{(\lambda_+(\bk)-\Omega)^2-(\lambda_-(\bk)-\Omega)^2}\\
    &-\frac{(W_0)_{22}(\bx,\bk,0)(\lambda_-(\bk)-\Omega)^2\left[(\lambda_+(\bk)-\Omega)^2+\coup^2\rho_0\right]}{\coup^2\rho_0\left[(\lambda_+(\bk)-\Omega)^2-(\lambda_-(\bk)-\Omega)^2\right]} \ ,\\
    \nonumber
\modeii(\bx,\bk,0)&=\frac{(W_0)_{11}(\bx,\bk,0)\left[(\lambda_-(\bk)-\Omega)^2+\coup^2\rho_0\right]}{(\lambda_-(\bk)-\Omega)^2-(\lambda_+(\bk)-\Omega)^2}\\
    &-\frac{(W_0)_{22}(\bx,\bk,0)(\lambda_+(\bk)-\Omega)^2\left[(\lambda_-(\bk)-\Omega)^2+\coup^2\rho_0\right]}{\coup^2\rho_0\left[(\lambda_-(\bk)-\Omega)^2-(\lambda_+(\bk)-\Omega)^2\right]} \ .
\label{eq:f12b}
\end{align}
The corresponding initial conditions for $u_\pm(\bx,\vert\bk\vert,t)$ are then given by
\begin{align}
    \nonumber u_+(\bx,\vert\bk\vert,0)&=\int d\hbk\frac{(W_0)_{11}(\bx,\bk,0)\left[(\lambda_+(\bk)-\Omega)^2+\coup^2\rho_0\right]}{(\lambda_+(\bk)-\Omega)^2-(\lambda_-(\bk)-\Omega)^2}\\
    &-\frac{(W_0)_{22}(\bx,\bk,0){(\lambda_-(\bk)-\Omega)^2}\left[(\lambda_+(\bk)-\Omega)^2+\coup^2\rho_0\right]}{\coup^2\rho_0\left[(\lambda_+(\bk)-\Omega)^2-(\lambda_-(\bk)-\Omega)^2\right]} \ , \\
\nonumber 
u_-(\bx,\vert\bk\vert,0)&=\int d\hbk\frac{(W_0)_{11}(\bx,\bk,0)\left[(\lambda_-(\bk)-\Omega)^2+\coup^2\rho_0\right]}{(\lambda_-(\bk)-\Omega)^2-(\lambda_+(\bk)-\Omega)^2}\\
    &-\frac{(W_0)_{22}(\bx,\bk,0)(\lambda_+(\bk)-\Omega)^2\left[(\lambda_-(\bk)-\Omega)^2+\coup^2\rho_0\right]}{\coup^2\rho_0\left[(\lambda_-(\bk)-\Omega)^2-(\lambda_+(\bk)-\Omega)^2\right]} \ .
\end{align}

\begin{figure}[t]
\begin{center}
\includegraphics[width=4.in]{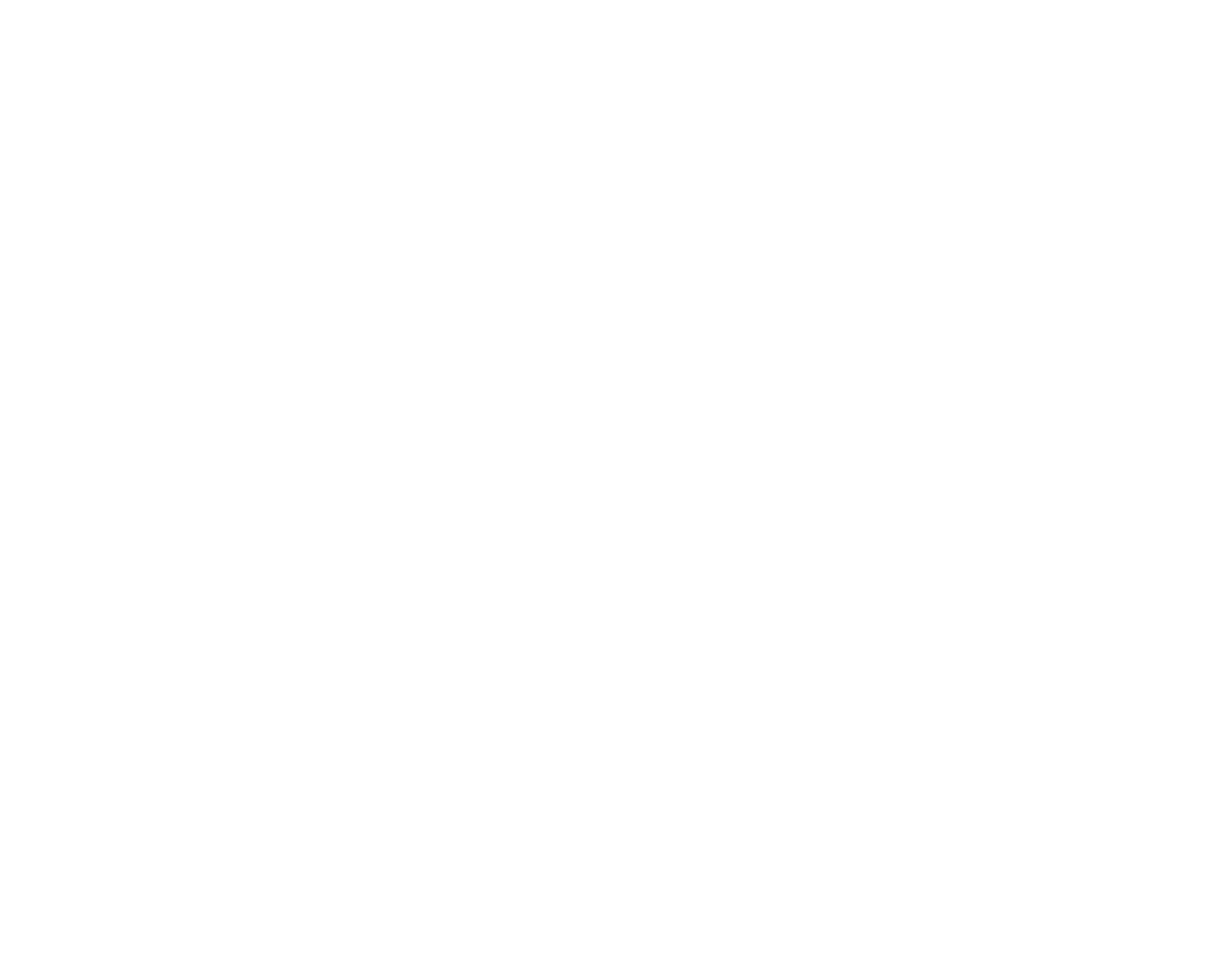}
\end{center}
\caption{\label{fig:fieldatom} Time dependence of the field and atomic probability densities in a random medium with $|\bx|=3l_s$.}
\end{figure}

\begin{figure}[t]
\begin{center}
\includegraphics[width=4.in]{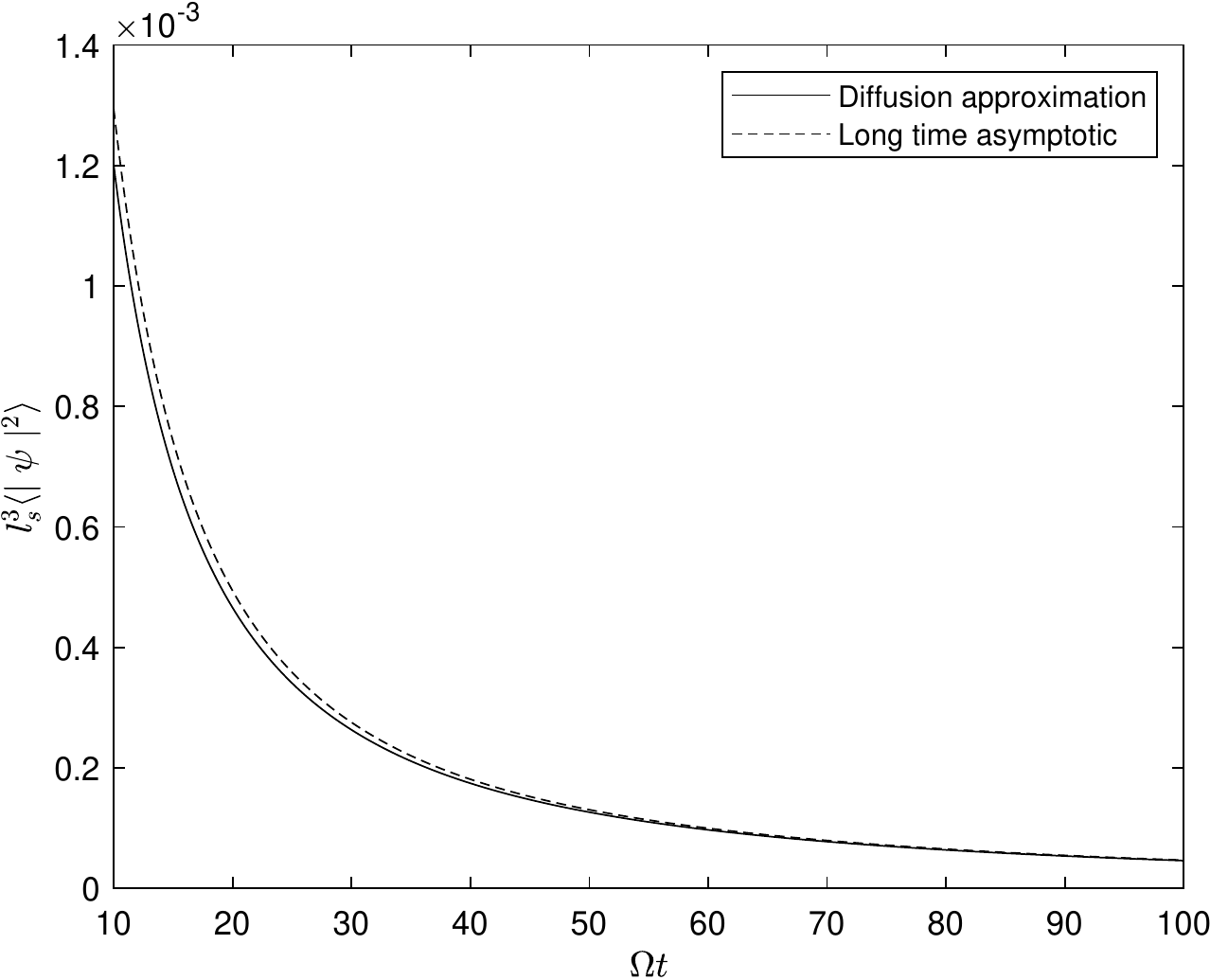}
\end{center}
\caption{\label{fig:asymptotic1}Long-time behavior of the field probability density when $\vert\bx\vert = l_s$.}
\end{figure}

We suppose that the atoms are initially excited near the origin in a volume of linear dimensions $l_s$ and that there are no photons present in the field. We thus impose the following initial conditions on the amplitudes:
\begin{align}
\sqrt{\rho_0}a(\bx,0) &=\left(\frac{1}{\pi l_s^2}\right)^{3/4}e^{-\vert\bx\vert^2/2l_s^2} \ , \\
\psi(\bx,0) &=0 \ .
\end{align}
The initial conditions inherited by $u_{\pm}$ are then given by
\begin{align}
\label{eq:g3a}
 u_{+}(\bx,\vert\bk\vert,0) &=\frac{4}{\pi^2}\frac{(\lambda_+(\bk)-\Omega)^2+\coup^2\rho_0}{(\lambda_-(\bk)-\Omega)^2-(\lambda_+(\bk)-\Omega)^2}\frac{(\lambda_-(\bk)-\Omega)^2}{\coup^2\rho_0 }e^{-l_s^2 \vert\bk\vert^2}e^{-\vert\bx\vert^2/l_s^2} \ , \\
    u_{-}(\bx,\vert\bk\vert,0) &=\frac{4}{\pi^2}\frac{(\lambda_-(\bk)-\Omega)^2+\coup^2\rho_0}{(\lambda_+(\bk)-\Omega)^2-(\lambda_-(\bk)-\Omega)^2}\frac{(\lambda_+(\bk)-\Omega)^2}{\coup^2\rho_0}e^{-l_s^2 \vert\bk\vert^2}e^{-\vert\bx\vert^2/l_s^2} \ .
\label{eq:g3b}
\end{align}
Using (\ref{soln}), we find that the solutions to the diffusion equations (\ref{eq:g2}) with initial conditions (\ref{eq:g3a}) and (\ref{eq:g3b}) are given by

\begin{align}
\nonumber u_{+}(\bx,\vert\bk\vert,t) &=\frac{4}{\pi^2}\frac{(\lambda_+(\bk)-\Omega)^2+\coup\rho_0 }{(\lambda_-(\bk)-\Omega)^2-(\lambda_+(\bk)-\Omega)^2}\frac{(\lambda_-(\bk)-\Omega)^2}{\coup^2\rho_0}e^{-l_s^2
    \vert\bk\vert^2}\\
    &\times\left(\frac{l_s^2}{l_s^2+4tD_+(\vert\bk\vert)}\right)^{3/2}e^{-\vert\bx\vert^2/(l_s^2+4tD_+(\vert\bk\vert))} \ , \\
     \nonumber u_{-}(\bx,\vert\bk\vert,t) &=\frac{4}{\pi^2}\frac{(\lambda_-(\bk)-\Omega)^2+\coup^2\rho_0 }{(\lambda_+(\bk)-\Omega)^2-(\lambda_-(\bk)-\Omega)^2}\frac{(\lambda_+(\bk)-\Omega)^2}{\coup^2\rho_0}e^{-l_s^2 \vert\bk\vert^2} \\
     &\times\left(\frac{l_s^2}{l_s^2+4tD_-(\vert\bk\vert)}\right)^{3/2}e^{-\vert\bx\vert^2/(l_s^2+4tD_-(\vert\bk\vert))} \ .
\end{align}
Using (\ref{eq:f11}), we see that the average probability densities are given by the formulas
\begin{align}
\label{eq:g4}
     \nonumber
&\langle\vert \psi(\bx,t)\vert^2\rangle\\
     \nonumber&=\frac{4}{\coup^2\rho_0 \pi^2} \int_0^{\infty} dk\,k^2 e^{-l_s^2
    k^2}\left[\frac{[(\lambda_-(k)-\Omega)(\lambda_+(k)-\Omega)]^2}{(\lambda_-(k)-\Omega)^2-(\lambda_+(k)-\Omega)^2}\left(\frac{l_s^2}{l_s^2+4tD_+(k)}\right)^{3/2}
    e^{-\vert\bx\vert^2/(l_s^2+4tD_+(k)}\right. \\
    & +\left.\frac{[(\lambda_+(k)-\Omega)(\lambda_-(k)-\Omega)]^2}{(\lambda_+(k)-\Omega)^2-(\lambda_-(k)-\Omega)^2}\left(\frac{l_s^2}{l_s^2+4tD_-(k)}\right)^{3/2}e^{-\vert\bx\vert^2/(l_s^2+4tD_-(k))}\right] \ ,
\end{align}
\begin{align}\label{eq:g5}
\nonumber
&\rho_0\langle\vert a(\bx,t)\vert^2\rangle\\
    \nonumber&=\frac{4}{\pi^2} \int_0^\infty dk\,k^2 e^{-l_s^2
    k^2} \left[\frac{(\lambda_-(k)-\Omega)^2}{(\lambda_-(k)-\Omega)^2-(\lambda_+(k)-\Omega)^2}\left(\frac{l_s^2}{l_s^2+4tD_+(k)}\right)^{3/2}e^{-\vert\bx\vert^2/(l_s^2+4tD_+(k))}\right.\\ 
    &+\left.\frac{(\lambda_+(k)-\Omega)^2}{(\lambda_+(k)-\Omega)^2-(\lambda_-(k)-\Omega)^2}\left(\frac{l_s^2}{l_s^2+4tD_-(k)}\right)^{3/2}e^{-\vert\bx\vert^2/(l_s^2+4tD_-(k))}\right].
\end{align}

At long times, we find that $\langle\vert a\vert^2\rangle$ and $\langle\vert\psi\vert^2\rangle$ decay algebraically according to
\begin{align}
    \rho_0\langle\vert a(\bx,t)\vert^2\rangle &= \frac{C_1}{t^{3/2}}-\frac{C_2}{t^{5/2}}\vert\bx\vert^2 \label{eq:g6} \ ,\\
    \langle\vert\psi(\bx,t)\vert^2\rangle &= \frac{C_3}{t^{3/2}} -\frac{C_4}{t^{5/2}}\vert\bx\vert^2 \label{eq:g7}
\end{align}
where the $C_i$ are given by
\begin{align}
\nonumber C_1 &=  \frac{4}{\pi^2} \int_0^\infty dk\,\left(\frac{k^2 e^{-l_s^2
    k^2}}{(\lambda_-(k)-\Omega)^2-(\lambda_+(k)-\Omega)^2}\right)\\
    &\times\left[(\lambda_-(k)-\Omega)^2\left(\frac{l_s^2}{4D_+(k)}\right)^{3/2}-(\lambda_+(k)-\Omega)^2\left(\frac{l_s^2}{4D_-(k)}\right)^{3/2}\right]\ ,\\
\nonumber C_2 &= \frac{4}{l_s^2 \pi^2} \int_0^\infty dk\,\left(\frac{k^2 e^{-l_s^2
    k^2}}{(\lambda_-(k)-\Omega)^2-(\lambda_+(k)-\Omega)^2}\right)\\
    &\times\left[(\lambda_-(k)-\Omega)^2\left(\frac{l_s^2}{4D_+(k)}\right)^{5/2}-(\lambda_+(k)-\Omega)^2\left(\frac{l_s^2}{4D_-(k)}\right)^{5/2}\right] \, ,\\
\nonumber C_3 &=  \frac{4}{\coup^2\rho_0\pi^2} \int_0^\infty dk\,k^2 e^{-l_s^2
    k^2}\left(\frac{(\lambda_-(k)-\Omega)^2(\lambda_+(k)-\Omega)^2}{(\lambda_-(k)-\Omega)^2-(\lambda_+(k)-\Omega)^2}\right)\\
    &\times\left[\left(\frac{l_s^2}{4D_+(k)}\right)^{3/2}-\left(\frac{l_s^2}{4D_-(k)}\right)^{3/2}\right]\ ,\\
\nonumber C_4 &= \frac{4}{l_s^2\coup^2\rho_0\pi^2} \int_0^\infty dk\,k^2 e^{-l_s^2
    k^2}\left(\frac{(\lambda_-(k)-\Omega)^2(\lambda_+(k)-\Omega)^2}{(\lambda_-(k)-\Omega)^2-(\lambda_+(k)-\Omega)^2}\right)\\
    &\times\left[\left(\frac{l_s^2}{4D_+(k)}\right)^{5/2}-\left(\frac{l_s^2}{4D_-(k)}\right)^{5/2}\right].
\end{align}

\begin{figure}[t]
\begin{center}
\includegraphics[width=4.in]{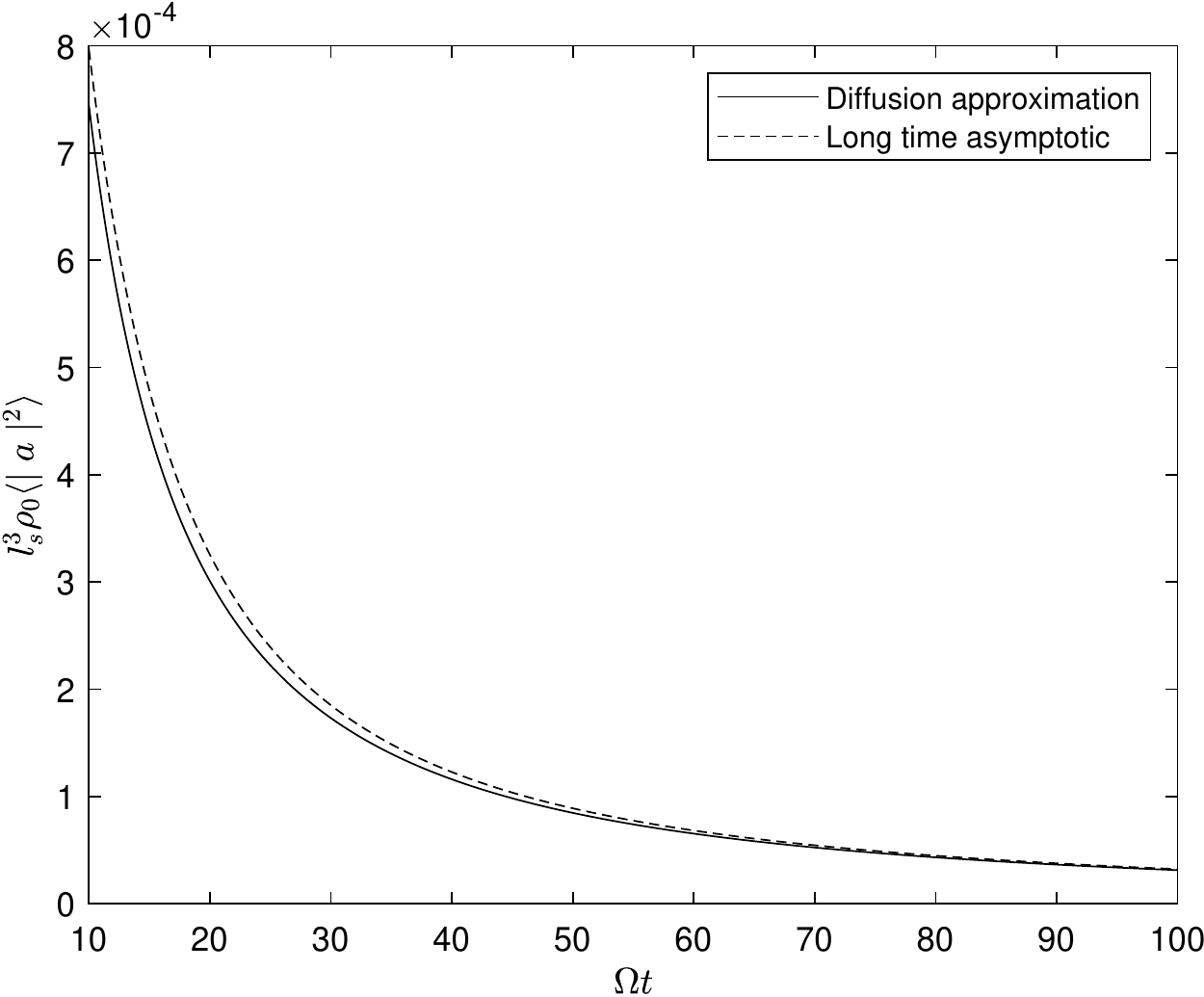}
\end{center}
\caption{\label{fig:asymptotic2}Long-time behavior of the atomic probability density when $\vert\bx\vert = l_s$.}
\end{figure}

To illustrate the above results, we consider isotropic scattering with $A=1/(4\pi)$, and put the dimensionless quantities $\Omega l_s/c=\rho_0(g/\Omega)^2 =1$. Figure~\ref{fig:multiplepositions} shows the time dependence of $\rho_0\langle\vert a(\bx,t)\vert^2\rangle$ for different values of $|\bx|$. As may be expected, $\rho_0\langle\vert a\vert^2\rangle$ decays faster at larger distances away from the initial volume of excitation. 
In Figure~\ref{fig:fieldatom} we plot the time dependence of $\langle\vert\psi\vert^2\rangle$ and
$\rho_0\langle\vert a\vert^2\rangle$ for $|\bx|=3l_s$. We note that the negative values of these quantities for small times are due to the breakdown of the diffusion approximation. In Figures~\ref{fig:asymptotic1}  and~\ref{fig:asymptotic2} we compare (\ref{eq:g4}) and~(\ref{eq:g5}) with the asymptotic formulas (\ref{eq:g6}) and (\ref{eq:g7}). There is good agreement at long times.


\section{Discussion}
\label{discussion}
We have investigated the problem of cooperative spontaneous emission in random media. Our main results are kinetic equations that govern the behavior of the one-photon and atomic probability densities. Several topics for further research are apparent. An alternative derivation of (\ref{RTE_modes}) may be possible using diagrammatic perturbation theory rather than multiscale asymptotic analysis. This is the case for the classical theory of wave propagation in random media,  where a comparative exposition of the two approaches has been presented in~\cite{Caze_2015}. It would also be of interest to examine the transport of two-photon states in random media. Here the evolution  of the entanglement of an initially entangled state is of particular importance, especially in applications to communications and imaging. Finally, it would be of interest to extend our results to polariton transport in random media consisting of atoms embedded in a dielectric. In this setting, the systems of equations (\ref{eq_psi}) and (\ref{eq_a}) is no longer nonlocal.

\appendix

\section{Derivation of (\ref{eq:b4}) and (\ref{eq:b5})}
Here we derive the system of equations (\ref{eq:b4}). To proceed, we compute both sides of the Schrodinger equation $i\hbar \partial_t \vert\Psi\rangle = H\vert\Psi\rangle$. Making use of 
\begin{align}
\label{state}
\vert\Psi\rangle = \int d^3 x \left[\psi(\bx,t)\phi^{\dagger}(\bx) + \rho(\bx)a(\bx,t)\sigma^{\dagger}(\bx)\right]\vert 0\rangle \ ,
\end{align}
we find that the left hand side is given by 
\begin{align}
 i\hbar\partial_t\vert\Psi\rangle=\int d^3 x \left[i\hbar\partial_t\psi(\bx,t)\phi^{\dagger}(\bx) +i\hbar\partial_t a(\bx,t)\rho(\bx)\sigma^{\dagger}(\bx)\right]\vert 0\rangle \ .
\end{align}
Next, using the definition (\ref{Hamiltonian}) of the Hamiltonian $H$ and the commutation relations (\ref{commutation}) and (\ref{anticommutation}), the right hand side becomes
\begin{align}
\nonumber 
H\vert\Psi\rangle  =
\int d^3 x \big[&\hbar c(-\Delta)^{1/2}\psi(\bx,t) \phi^{\dagger}(\bx)+\hbar g\rho(\bx)\psi(\bx,t)\sigma^{\dagger}(\bx) 
 + \hbar g\rho(\bx)a(\bx,t)\phi^{\dagger}(\bx) \\
&+\hbar\Omega\rho(\bx)a(\bx,t)\sigma^{\dagger}(\bx)\big]\vert 0\rangle \ .
\end{align}
It follows that 
\begin{align}
\langle 0\vert\phi(\bx) i\hbar\partial_t\vert\Psi\rangle &= i\hbar\partial_t\psi(\bx,t) \ ,\\
\langle 0\vert\phi(\bx) H\vert\Psi\rangle &=\hbar c((-\Delta)^{1/2}\psi)(\bx,t)+\hbar g\rho(\bx)a(\bx,t) \ .
\end{align}
Likewise
\begin{align}
    \langle 0\vert\sigma(\bx)\rho(\bx)i\hbar\partial_t\vert\Psi\rangle &= i\hbar\partial_t a(\bx,t) \ , \\
    \langle 0\vert\sigma(\bx)\rho(\bx)H\vert\Psi\rangle &=\hbar g\rho(\bx)\psi(\bx,t)+\hbar\Omega\rho(\bx)a(\bx,t) \ .
\end{align}
We thus obtain
\begin{align}
i\partial_t\psi & = c(-\Delta)^{1/2}\psi + \coup\rho(\bx) a \ , \\
i\rho(\bx)\partial_t a & = \coup\rho(\bx)\psi + \Omega\rho(\bx) a \ ,
\end{align}
which are (\ref{eq:b4}) and (\ref{eq:b5}).

\section{Derivation of Eq.~(\ref{eq:d6})}
We proceed from (\ref{eq:d5}):
\begin{align}
\epsilon c\left[(-\Delta_{\bx_1})^{1/2}-(-\Delta_{\bx_2})^{1/2}\right]\Phi_{\epsilon}(\bx_1,\bx_2)+\sqrt{\epsilon}\frac{\rho_0 g^2}{\omega-\Omega}\left[\eta(\bx_1/\epsilon)-\eta(\bx_2/\epsilon)\right]\Phi_{\epsilon}(\bx_1,\bx_2)=0
\end{align}
and make the change of variables 
\begin{align}
    \bx_1 &=\bx-\epsilon\bx'/2 \ ,\\
    \bx_2 &=\bx+\epsilon\bx'/2 \ .
\end{align}
We then Fourier transform the result with respect to $\bx'$. 
The first term becomes
\begin{align}
     \nonumber&\epsilon c\int\frac{\dd^3 x'}{(2\pi)^3}\,e^{-i\bk\cdot\bx'}\left[(-\Delta_{\bx_1})^{1/2}-(-\Delta_{\bx_2})^{1/2}\right]\Phi_{\epsilon}(\bx-\epsilon\bx'/2,\bx+\epsilon\bx'/2)\\
     \nonumber=&\epsilon c\int\,\frac{\dd^3 x'}{(2\pi)^3}\frac{\dd^3 q_1}{(2\pi)^3}\frac{\dd^3 q_2}{(2\pi)^3}e^{-i\bk\cdot\bx'}e^{i\bq_1\cdot(\bx-\epsilon\bx'/2)}e^{-i\bq_2\cdot(\bx+\epsilon\bx'/2)}\left[\vert\bq_1\vert-\vert\bq_2\vert\right]\tilde{\Phi}_{\epsilon}(\bq_1,\bq_2)\\
     =& c\int\frac{\dd^3 q}{(2\pi)^3}e^{i\bq\cdot\bx}\left[\vert -\bk+\epsilon\bq/2\vert-\vert\bk+\epsilon\bq/2\vert\right]\tilde{W}_{\epsilon}(\bq,\bk) \ .
\end{align}
Continuing with the second term we have
\begin{align}
    &\sqrt{\epsilon}\frac{\rho_0 g^2}{\omega-\Omega}\int\frac{\dd^3 x'}{(2\pi)^3}\,e^{-i\bk\cdot\bx'}\,\eta(\bx/\epsilon-\bx'/2)\Phi_{\epsilon}(\bx-\epsilon\bx'/2,\bx+\epsilon\bx'/2)\\
    =&\sqrt{\epsilon}\frac{\rho_0 g^2}{\omega-\Omega}\int\frac{\dd^3 q}{(2\pi)^3}\,e^{i\bq\cdot\bx/\epsilon}\,\tilde{\eta}(\bq)W_{\epsilon}(\bx,\bk+\bq/2) \ .
\end{align}
The third term follows similarly:
\begin{align}
    &\sqrt{\epsilon}\frac{\rho_0 g^2}{\omega-\Omega}\int\frac{\dd^3 x'}{(2\pi)^3}\,e^{-i\bk\cdot\bx'}\,\eta(\bx/\epsilon+\bx'/2)\Phi_{\epsilon}(\bx-\epsilon\bx'/2,\bx+\epsilon\bx'/2)\\
    =&\sqrt{\epsilon}\frac{\rho_0 g^2}{\omega-\Omega}\int\frac{\dd^3 q}{(2\pi)^3}\,e^{i\bq\cdot\bx/\epsilon}\,\tilde{\eta}(\bq)W_{\epsilon}(\bx,\bk-\bq/2) \ .
\end{align}
Combining the above yields (\ref{eq:d6}):
\begin{align} 
\int\frac{d^3 q}{(2\pi)^3}e^{i\bq\cdot\bx}\left[\vert -\bk+\epsilon\bq/2\vert-\vert\bk+\epsilon\bq/2\vert\right]\tilde{W}_{\epsilon}(\bq,\bk)+ \sqrt{\epsilon}LW_{\epsilon}(\bx,\bk)=0 \ ,
\end{align}
where 
\begin{align}
LW_{\epsilon}(\bx,\bk)=k_0\int\frac{d^3 q}{(2\pi)^3}e^{i\bq\cdot\bx/\epsilon}\tilde{\eta}(\bq)\left[W_{\epsilon}(\bx,\bk+\bq/2)-W_{\epsilon}(\bx,\bk-\bq/2)\right] \ .
\end{align}

\section{Derivation of Eq.~(\ref{eq:d13})}
We evaluate the second term in ~(\ref{eq:d12}) using (\ref{eq:d10})
\begin{align}
 &k_0\int\frac{d^3 q}{(2\pi)^3}e^{i\bq\cdot\bX}\langle\tilde{\eta}(\bq)\left[W_1(\bx,\bX,\bk+\bq/2)-W_1(\bx,\bX,\bk-\bq/2)\right]\rangle\\
    =\nonumber&k_0^2\int\frac{d^3 q}{(2\pi)^3}\frac{d^3 Q}{(2\pi)^3}e^{i\bq\cdot\bX+i\bQ\cdot\bX}\langle\tilde{\eta}(\bq)\left[\tilde{\eta}(\bQ)\frac{[W_0(\bx,\bk+\bq/2+\bQ/2)-W_0(\bx,\bk+\bq/2-\bQ/2)]}{\vert\bk+\bq/2+\bQ/2\vert-\vert -\bk-\bq/2+\bQ/2\vert-i\theta}\right.\\
    &-\left.\tilde{\eta}(\bQ)\frac{[W_0(\bx,\bk-\bq/2+\bQ/2)-W_0(\bx,\bk-\bq/2-\bQ/2)]}{\vert\bk-\bq/2+\bQ/2\vert-\vert -\bk+\bq/2+\bQ/2\vert-i\theta}\right]\rangle\\
    =&k_0^2\int\frac{d^3 q}{(2\pi)^3}\tilde{C}(\bq-\bk)\left(W_0(\bx,\bk)-W_0(\bx,\bq)\right)\left[\frac{1}{\vert\bk\vert-\vert \bq\vert-i\theta}-\frac{1}{\vert\bk\vert-\vert\bq\vert+i\theta}\right] \ ,
\end{align}
where we have made use of the relation
\begin{align}
    \langle\tilde{\eta}(\bq)\tilde{\eta}(\bQ)\rangle = (2\pi)^3\tilde{C}(\bq)\delta(\bq+\bQ) \ .
\end{align}
Finally we put $\theta\to 0$ and use the identity
\begin{align}
\label{identity}
    \lim_{\theta\to 0}\left(\frac{1}{x-i\theta}-\frac{1}{x+i\theta}\right)= 2\pi i\delta(x)\ ,
\end{align}
to obtain
\begin{align}
    k_0^2\int\frac{d^3 q}{(2\pi)^2}\tilde{C}(\bq-\bk)\left(W_0(\bx,\bk)-W_0(\bx,\bq)\right)\delta(\vert\bk\vert-\vert\bq\vert) \ ,
\end{align}
which yields (\ref{eq:d13}).

\section{Derivation of Eq.~(\ref{eq:f1})}
We begin with the equation satisfied by $\Phi_{\epsilon}(\bx_1,\bx_2)=\bu_{\epsilon}(\bx_1)\bu_{\epsilon}^*(\bx_2)$:
\begin{align}
    \nonumber\epsilon i\partial_t\Phi_{\epsilon}&= A_{\epsilon}(\bx_1)\Phi_{\epsilon}+\sqrt{\epsilon}\coup\sqrt{\rho_0}\eta(\bx_1/\epsilon)K\Phi_{\epsilon}\\
    &-\Phi_{\epsilon}A_{\epsilon}(\bx_2)+\sqrt{\epsilon}\coup\sqrt{\rho_0}\eta(\bx_2/\epsilon)\Phi_{\epsilon}K^{T}.
\end{align}
Next we make the change of variables
\begin{align}
    \bx_1&=\bx-\epsilon\bx'/2 \ , \\
    \bx_2&=\bx+\epsilon\bx'/2 \ ,
\end{align}
and Fourier transform the result with respect to $\bx'$, which leads to
\begin{align} 
\label{eq:AE1}
\nonumber
\epsilon i\partial_t W_{\epsilon}(\bx,\bk,t)&= \int d^3 x'e^{-i\bk\cdot\bx'}\big[A_{\epsilon}(\bx-\epsilon\bx'/2)\Phi_{\epsilon}(\bx-\epsilon\bx'/2,\bx+\epsilon\bx'/2)
\\
\nonumber
&-\Phi_{\epsilon}(\bx-\epsilon\bx'/2,\bx+\epsilon\bx'/2)A_{\epsilon}(\bx+\epsilon\bx'/2)\big]\\
     &\nonumber+\sqrt{\epsilon}\coup\sqrt{\rho_0}\int d^3 x'e^{-i\bk\cdot\bx'}\eta(\bx/\epsilon-\bx'/2)K\Phi_{\epsilon}(\bx-\epsilon\bx'/2,\bx+\epsilon\bx'/2)\\
     &+\sqrt{\epsilon}\coup\sqrt{\rho_0}\int d^3 x'e^{-i\bk\cdot\bx'}\eta(\bx/\epsilon+\bx'/2)\Phi_{\epsilon}(\bx-\epsilon\bx'/2,\bx+\epsilon\bx'/2)K^{T}.
\end{align}
The first term on the right hand side of (\ref{eq:AE1}) becomes
\begin{align}
    &\nonumber\int d^3 x'e^{-i\bk\cdot\bx'}\left[A_{\epsilon}(\bx-\epsilon\bx'/2)\Phi_{\epsilon}(\bx-\epsilon\bx'/2,\bx+\epsilon\bx'/2)-\Phi_{\epsilon}(\bx-\epsilon\bx'/2,\bx+\epsilon\bx'/2)A_{\epsilon}(\bx+\epsilon\bx'/2)\right]\\
    =&\nonumber\int d^3 x'\frac{d^3 q_1}{(2\pi)^3}\frac{d^3 q_2}{(2\pi)^3}e^{-i\bk\cdot\bx'+i\bq_1\cdot(\bx-\epsilon\bx'/2)-i\bq_2\cdot(\bx+\epsilon\bx'/2)}\left[\tilde{A}_{\epsilon}(\bq_1)\tilde{\Phi}_{\epsilon}(\bq_1,\bq_2)-\tilde{\Phi}_{\epsilon}(\bq_1,\bq_2)\tilde{A}_{\epsilon}(\bq_2)\right]\\
     =&\int\frac{d^3 q}{(2\pi)^3}e^{i\bx\cdot\bq}\left[\tilde{A}_{\epsilon}(\bk/\epsilon-\bq/2)\tilde{W}_{\epsilon}(\bq,\bk)-\tilde{W}_{\epsilon}(\bq,\bk)\tilde{A}_{\epsilon}(\bk/\epsilon+\bq/2)\right] \ .
\end{align}
The second term is seen to be
\begin{align}
    &\nonumber\sqrt{\epsilon}\coup\sqrt{\rho_0}\int d^3 x'e^{-i\bk\cdot\bx'}\eta(\bx/\epsilon-\bx'/2)K\Phi_{\epsilon}(\bx-\epsilon\bx'/2,\bx+\epsilon\bx'/2)\\
    =&\sqrt{\epsilon}\coup\sqrt{\rho_0}\int \frac{d^3 q}{(2\pi)^3} e^{i\bq\cdot\bx/\epsilon}\tilde{\eta}(\bq)KW_{\epsilon}(\bx,\bk+\bq/2)  \ .
\end{align}
The third term is handled similarly:
\begin{align}
      &\nonumber\sqrt{\epsilon}\coup\sqrt{\rho_0}\int d^3 x'e^{-i\bk\cdot\bx'}\eta(\bx/\epsilon+\bx'/2)\Phi_{\epsilon}(\bx-\epsilon\bx'/2,\bx+\epsilon\bx'/2)K^{T}\\
    =&\sqrt{\epsilon}\coup\sqrt{\rho_0}\int \frac{d^3 q}{(2\pi)^3} e^{i\bq\cdot\bx/\epsilon}\tilde{\eta}(\bq)W_{\epsilon}(\bx,\bk-\bq/2) \ .
\end{align}
Putting the above together yields (\ref{eq:f1}).

\section{Derivation of Eq.~(\ref{RTE_modes})}

The first two terms on the left hand side of (\ref{RTE_modes}) are easily obtained. The remaining terms come from considering
\begin{align}
    \langle\bb_+^{T}(\bk)\coup\sqrt{\rho_0}\int\frac{d^3 q}{(2\pi)^3}e^{i\bq\cdot\bX}\tilde{\eta}(\bq)\left[KW_{1}(\bx,\bX,\bk+\bq/2,t)-W_{1}(\bx,\bX,\bk-\bq/2,t)K^{T}\right]\bb_+(\bk)\rangle
\end{align}
The first term above is given by
\begin{align}
\label{first}
    &\nonumber\langle\bb_+^{T}(\bk)\int\frac{d^3 q}{(2\pi)^3}e^{i\bq\cdot\bX}\tilde{\eta}(\bq)KW_{1}(\bx,\bX,\bk+\bq/2,t)\bb_+(\bk)\rangle\\
    =&\nonumber\int\frac{d^3 q}{(2\pi)^3}\frac{d^3 Q}{(2\pi)^3}e^{i\bq\cdot\bX+i\bQ\cdot\bX}\langle\bb_+^{T}(\bk)\tilde{\eta}(\bq)K\sum_{m,n}w_{mn}(\bx,\bQ,\bk+\bq/2,t)\bb_m(\bk+\bq/2-\bQ/2)\\
    \nonumber&\times\bb_n^T(\bk+\bq/2+\bQ/2)\bb_+(\bk)\rangle\\
    =&\int\frac{d^3 q}{(2\pi)^3}\tilde{C}(\bk-\bq)\sum_{m}\frac{(\coup^2\rho_0)^{3/2}(\lambda_+(\bk)-\Omega)((\lambda_+(\bk)-\Omega)\modem(\bx,\bq,t)-(\lambda_m(\bq)-\Omega)\modei(\bx,\bk,t)}{((\lambda_m(\bq)-\Omega)^2+\coup^2\rho_0)((\lambda_+(\bk)-\Omega)^2+\coup^2\rho_0)(\lambda_{m}(\bq)-\lambda_{+}(\bk)+i\theta)} \ .
\end{align}
The second term becomes
\begin{align}
\label{second}
    &\nonumber\langle\bb_+^{T}(\bk)\int\frac{d^3 q}{(2\pi)^3}e^{i\bq\cdot\bX}\tilde{\eta}(\bq)W_{1}(\bx,\bX,\bk-\bq/2,t)K^{T}\bb_+(\bk)\rangle\\
    =&\nonumber\int\frac{d^3 q}{(2\pi)^3}\frac{d^3 Q}{(2\pi)^3}e^{i\bq\cdot\bX+i\bQ\cdot\bX}\langle\bb_+^{T}(\bk)\tilde{\eta}(\bq)\sum_{m,n}w_{mn}(\bx,\bQ,\bk-\bq/2,t)\bb_m(\bk-\bq/2-\bQ/2)\\
    \nonumber&\times\bb_n^T(\bk-\bq/2+\bQ/2)K^{T}\bb_+(\bk)\rangle\\
    =&\int\frac{d^3 q}{(2\pi)^3}\tilde{C}(\bk-\bq)\sum_{n}\frac{(\coup^2\rho_0)^{3/2}(\lambda_+(\bk)-\Omega)((\lambda_n(\bq)-\Omega)\modei(\bx,\bk,t)-(\lambda_+(\bk)-\Omega)\moden(\bx,\bq,t)}{((\lambda_+(\bk)-\Omega)^2+\coup^2\rho_0)((\lambda_n(\bq)-\Omega)^2+\coup^2\rho_0)(\lambda_{+}(\bk)-\lambda_{n}(\bq)+i\theta))} \ .
\end{align}
Subtracting (\ref{first}) and (\ref{second}), letting $\theta\to0$ and using (\ref{identity}) yields
\begin{align}
\nonumber
&\langle\bb_+^{T}(\bk)\coup\sqrt{\rho_0}\int\frac{d^3 q}{(2\pi)^3}e^{i\bq\cdot\bX}\tilde{\eta}(\bq)\left[KW_{1}(\bx,\bX,\bk+\bq/2,t)-W_{1}(\bx,\bX,\bk-\bq/2,t)K^{T}\right]\bb_+(\bk)\rangle\\
    =&\frac{2\pi(\coup^2\rho_0)^{2}(\lambda_+(\bk)-\Omega)^2}{(\lambda_+(\bq)-\Omega)^2+\coup^2\rho_0)^2}\int\frac{d^3 q}{(2\pi)^3}\tilde{C}(\bk-\bq)\delta(\lambda_{+}(\bq)-\lambda_{+}(\bk))\left[\modei(\bx,\bk,t)-\modei(\bx,\bq,t)\right] \ ,
\end{align}
where only the $m=+$ contribution is included. Putting everything together we see that $\modei$ satisfies the equation 
\begin{align} \label{eq:AE2}
    \nonumber\frac{1}{c}\partial_t\modei(\bx,\bk,t) &+\left(\frac{(\lambda_{+}(\bk)-\Omega)^2}{(\lambda_{+}(\bk)-\Omega)^2+\coup^2\rho_0}\right)\hbk\cdot\nabla_{\bx}\modei(\bx,\bk,t)\\
    \nonumber&+\left[\frac{2\pi (\coup^2\rho_0)^2(\lambda_+(\bk)-\Omega)^2}{c((\lambda_+(\bk)-\Omega)^2+\coup^2\rho_0)^2}\int\frac{d^3 q}{(2\pi)^3}\,\tilde{C}(\bk-\bq)\delta(\lambda_+(\bq)-\lambda_+(\bk))\right]\modei(\bx,\bk,t)\\
    &=\frac{2\pi (\coup^2\rho_0)^2(\lambda_+(\bk)-\Omega)^2}{c((\lambda_+(\bk)-\Omega)^2+\coup^2\rho_0)^2}\int\frac{d^3 q}{(2\pi)^3}\,\tilde{C}(\bk-\bq)\delta(\lambda_+(\bq)-\lambda_+(\bk))\modei(\br,\bq,t) \ .
\end{align}
The delta function $\delta(\lambda_+(\bq)-\lambda_+(\bk))$ can be expressed as 
\begin{align}
    \delta(\lambda_+(\bq)-\lambda_+(\bk)) 
    &=2\delta(\vert\bq\vert-\vert\bk\vert)\frac{\sqrt{(c\vert\bk\vert-\Omega)^2+4\coup^2\rho_0}}{c(\lambda_+(\bk)-\Omega)} \ .
\end{align}
Hence (\ref{eq:AE2}) becomes
\begin{align} 
     \nonumber
&\frac{1}{c}\partial_t\modei(\bx,\bk,t) +\left(\frac{(\lambda_{+}(\bk)-\Omega)^2}{(\lambda_{+}(\bk)-\Omega)^2+\coup^2\rho_0}\right)\hbk\cdot\nabla_{\bx}\modei(\bx,\bk,t)\\
    \nonumber&+\left[\frac{4\pi (\coup^2\rho_0)^2\vert\lambda_+(\bk)-\Omega\vert}{c^2((\lambda_+(\bk)-\Omega)^2+\coup^2\rho_0)^2}\sqrt{(c\vert\bk\vert-\Omega)^2+4\coup^2\rho_0}\vert\bk\vert^2\int\frac{d\hbk'}{(2\pi)^3}\,\tilde{C}(\vert\bk\vert(\hbk-\hbk'))\right]\modei(\bx,\bk,t)\\
    &=\frac{4\pi (\coup^2\rho_0)^2\vert\lambda_+(\bk)-\Omega\vert}{c^2((\lambda_+(\bk)-\Omega)^2+\coup^2\rho_0)^2}\sqrt{(c\vert\bk\vert-\Omega)^2+4\coup^2\rho_0}\vert\bk\vert^2\int\frac{d\hbk'}{(2\pi)^3}\,\tilde{C}(\vert\bk\vert(\hbk-\hbk'))\modei(\br,\bk',t) \ ,
\end{align}
which is (\ref{RTE_modes}).
The equation for $\modeii$ is derived in the same manner.

\section*{Acknowledgments}
This work was performed when the authors were members of the Department of Mathematics at University of Michigan.
We thank Jeremy Hoskins for valuable discussions. This work was supported in part by the NSF grant DMS-1912821 and the AFOSR grant FA9550-19-1-0320.

\end{document}